# Machine learning accelerates fuel cell life testing

Yanbin Zhao†1, Hao Liu†*1, Zhihua Deng2, Haoyi Jiang1, Zhenfei Ling1, Zhiyang Liu3, Xingkai Wang4, Tong Li1, Xiaoping Ouyang1

[1] State Key Laboratory of Fluid Power and Mechatronic Systems, School of Mechanical Engineering, Zhejiang University, Hangzhou, China.
[2] Energy Research Institute at NTU (ERI@N), Nanyang Technological University, Singapore.
[3] School of Control Science and Engineering, Zhejiang University, Hangzhou, China.
[4] School of Materials Science and Engineering, Tianjin University, Tianjin, China.
† These authors contributed equally to this work: Yanbin Zhao, Hao Liu.
* Corresponding author's email: haoliu7850052@zju.edu.cn (Hao Liu).

## Abstract

Accelerated life testing (ALT) can significantly reduce the economic, time, and labor costs of life testing in the process of equipment, device, and material research and development (R&D), and improve R&D efficiency. This paper proposes a performance characterization data prediction (PCDP) method and a life prediction-driven ALT (LP-ALT) method to accelerate the life test of polymer electrolyte membrane fuel cells (PEMFCs). The PCDP method can accurately predict different PCD using only four impedances (real and imaginary) corresponding to a high frequency and a medium frequency, greatly shortening the measurement time of offline PCD and reducing the difficulty of life testing. The test results on an open source life test dataset containing 42 PEMFCs show that compared with the determination coefficient ($R^2$) results of predicted aging indicators, including limiting current, total mass transport resistance, electrochemically active surface area, and crossover current, obtained based on the measured PCD, the $R^2$ results of predicted aging indicators based on the predicted PCD is only reduced by 0.04, 0.01, 0.05, and 0.06, respectively. Furthermore, the generalization of the PCDP method is verified by using an open-source life test dataset of a proton exchange membrane water electrolysis cell containing 28 times PCD measurement during approximately 1400 h. The LP-ALT method can shorten the life test time through early life prediction. Test results on the same open-source life test dataset of PEMFCs show that the acceleration ratio of the LP-ALT method can reach 30 times under the premise of ensuring that the minimum $R^2$ of the prediction results of different aging indicators, including limiting current, total mass transport resistance, and electrochemically active surface area, is not less than 0.89. Furthermore, the generalization of the LP-ALT method is verified by using an open-source life test dataset containing 24 capacitors. Combining the different performance characterization data predicted by the PCDP method and the life prediction of the LP-ALT method, the diagnosis and prognosis of PEMFCs and their components can be achieved.

## Introduction

Life testing (also known as stress testing or durability testing) [1, 2] plays an important role in the research and development (R&D) of equipment, devices and materials. Taking polymer electrolyte membrane fuel cells (PEMFCs) as an example, life testing is widely used in the R&D of new materials [3-5], new structures [6-8], new management strategies [9-11], etc. Accelerated life testing (ALT) can significantly reduce the cost of life testing and improve R&D efficiency [12]. At present, the life testing of PEMFCs is mainly accelerated by designing accelerated test conditions (ATCs) [13-17]. According to the study of Katayama et al. [18], after testing for about 75 hours under ATC, the aging state of the membrane electrode assembly (MEA) in PEMFCs is close to that of about 6000 hours operation on actual fuel cell vehicles, and the acceleration ratio can reach about 80 times. In addition to using ATC, life testing can also be accelerated by life prediction. Attia et al. [12] shortened the life test time of 224 lithium-ion (Li-ion) batteries from more than 500 days to 16 days through life prediction [19], greatly improving the development efficiency of fast-charging protocols and reducing life test costs. Although there are many studies on PEMFC life prediction methods, there is currently no life prediction-driven ALT (LP-ALT) method for PEMFCs.

During the life test of PEMFCs, reference performance tests (RPTs) [20, 21] (also known as check-up tests [22]) need to be carried out regularly to evaluate the aging state of PEMFCs. Performance characterization data (PCD) [23], such as polarization (I-V) curves, limiting current ($I_{lim}$) curves, electrochemical impedance spectroscopy (EIS), cyclic voltammetry (CV) curves, and linear sweep voltammetry (LSV) curves, are usually measured to analyze the aging state of PEMFCs and their components. For example, the electrochemically active surface area ($ECSA$) [24-26] extracted from the $CV$ curve can be utilized to characterize the aging state of the catalyst. The crossover current ($I_{\mathrm{cross}}$) parameter [27, 28] extracted from the LSV curve can be utilized to characterize the aging state of the proton exchange membrane. The total mass transport resistance ($R_{O2,total}$) [29-33] extracted from the $I_{lim}$ curve or $EIS$ can be utilized to characterize the aging state of the catalyst layer (CL) and gas diffusion layer (GDL). However, the commonly used PCD mentioned above all need to be measured offline, and some even need to rely on additional test equipment (such as EIS measurement) and complex test conditions (such as LSV curve measurement) [23, 34], which will not only increase the cost of life testing, but also increase the difficulty. Therefore, it is necessary to propose a method that can quickly obtain different PCD at a low cost.

This paper first proposes a method for PEMFC performance characterization data prediction (PCDP), which can achieve accurate prediction of different PCD, including EIS, I-V curve, CV curve, and LSV curve, by using only four impedances corresponding to a high frequency and a medium frequency. Moreover, accurate prediction of different aging indicators, including $R_{O2,total}$, $I_{lim}$, $ECSA$, and $I_{\mathrm{cross}}$, is achieved based on the predicted PCD. Then, this paper proposes a LP-ALT method for accelerating PEMFC life test, which mainly includes four steps: PCD collection, difference curve calculation, feature extraction, and life test prediction. Test results on the open-source life test dataset containing 42 PEMFCs [23] show that the proposed LP-ALT method can significantly reduce the cost of PEMFC life testing with a high acceleration ratio. Furthermore, the generalization of the proposed PCDP and LP-ALT methods are verified using an open-source life test dataset of a proton exchange membrane water electrolysis (PEMWE) [35] and an open-source life test dataset containing 24 capacitors [36], respectively.

The main contributions of this paper are: (1) The PCDP method and L P-ALT method for accelerating the life test of PEMFCs are proposed, which can significantly reduce the costs of life test (including economic cost, time cost, and labor cost) and improve the R&D efficiency. (2) The proposed PCDP and LP-ALT methods have good generalization performance, and can be promoted to accelerate the life test of other equipment, devices, and materials. (3) Combining different PCD predicted by the PCDP method and the life prediction of the LP-ALT method, the diagnosis and prognosis of PEMFCs and their components can be achieved.

## Results

### Datasets

The open-source life test dataset containing 42 PEMFCs (named as Dataset 1 in this paper) from Schneider et al. [23] is utilized to verify the proposed PCDP and LP-ALT methods. Dataset 1 contains a total of 42 independent PEMFC life tests and provides offline PCD at specific test stages, including I-V curves (from 0, 1k, 5k, 10k, and 30k cycles), EIS (from 0, 1k, 5k, 10k, and 30k cycles), CV curves (from 0, 10, 100, 1k, 3k, 5k, 10k, 20k, and 30k cycles), and LSV curves (from 0 and 30k cycles), as shown in **Supplementary Figure 1**. In each individual PEMFC life test, the composition and morphology parameters of the cathode catalyst layer were set differently, including platinum (Pt) load, ionomer to carbon (I/C) ratio, platinum on carbon (Pt/C) ratio, Pt alloy, ionomer equivalent weight (EW), and carbon support type parameters, as well as the ATCs used were different. The ATCs applied to all PEMFCs are rectangular potential cycles with the same period (6.5 s) and different upper potential limits (UPLs) and different relative humidity. The dwell time at UPLs (0.95 V or 1.15 V) and the low potential limit (LPL) (0.6 V) is 3 s, and the switching time between UPLs and LPL is 0.25 s. Each PEMFC was tested for 30,000 rectangular potential cycles, and the total life test time of the 42 PEMFCs exceeded 4000 h. For more details about Dataset 1, please refer to [23]. In this paper, since the three PEMFCs numbered 6, 33, and 42 in Dataset 1 did not have complete EIS data, they were eliminated. Then, we assign 11 PEMFCs to the test set (numbered as 3, 9, 12, 15, 18, 21, 24, 27, 30, 36, and 39 in [23]), and the remaining 28 PEMFCs to the training set.

The open-source life test dataset of a PEMWE (named Dataset 2 in this paper) from Rex et al. [35] is utilized to verify the generalization of the PCDP method. Dataset 2 contains the life test data of a PEMWE cell. The applied accelerated life test conditions are consists of 180 15-minute segments with varying current. Dataset 2 provides offline PCD measurements every 50 h. Specifically, in the offline PCD measurement phase, three I-V curve measurements are first performed, each measurement containing 38 constant current steps. Then, a full EIS measurement with a frequency range of 10 kHz to 0.1 Hz is performed. Next, an open circuit condition is performed for 1 h. Finally, three I-V curve measurements and a full EIS measurement are performed again. The PEMWE cell is test for approximately 1400 h, and the PCD measurements are repeated 28 times, as shown in **Supplementary Figure 2**. For more details about Dataset 2, please refer to [35]. In this paper, we assign the data obtained from 8 offline PCD measurements into the test set (numbered as 4th, 8th, 12th, 16th, 20th, 24th, and 28th in [35]), and the remaining 20 offline PCD measurements into the training set.

The open-source life test dataset containing 24 capacitors (named Dataset 3) from Renwick et al. [36] is utilized to verify the generalization of the LP-ALT method. Dataset 3 contains the life test data of 24 capacitors (the rated capacitance is 2200 μF, the maximum rated voltage is 10 V, the maximum rated current is 1 A, and the maximum operating temperature is 85 °C). Offline PCD, i.e., EIS, are provided for each capacitor at 72 different test stages from 0 h to 5105.5 h, as shown in **Supplementary Figure 3**. The applied accelerated life test conditions are electrical overstress accelerated test conditions, where the capacitors are charged and discharged with a 1 Hz square wave with an amplitude of 10 V, 12 V, or 14 V, and a load of 100 Ω. Each capacitor is tested for more than 5100 h, and the total test time of the 24 capacitors exceeded 122400 h. For more details about Dataset 3, please refer to [35]. In this paper, we assign 9 capacitors to the test set (numbered as ES10C1, ES10C2, ES10C3, ES12C1, ES12C2, ES10C3, ES14C1, ES14C2 and ES14C3 in [36]), and the remaining 15 capacitors to the training set.

### PCDP method

The proposed PCDP method consists of three steps, namely, impedances measurement, EIS prediction, and PCD prediction, as shown in **Figure 1**.

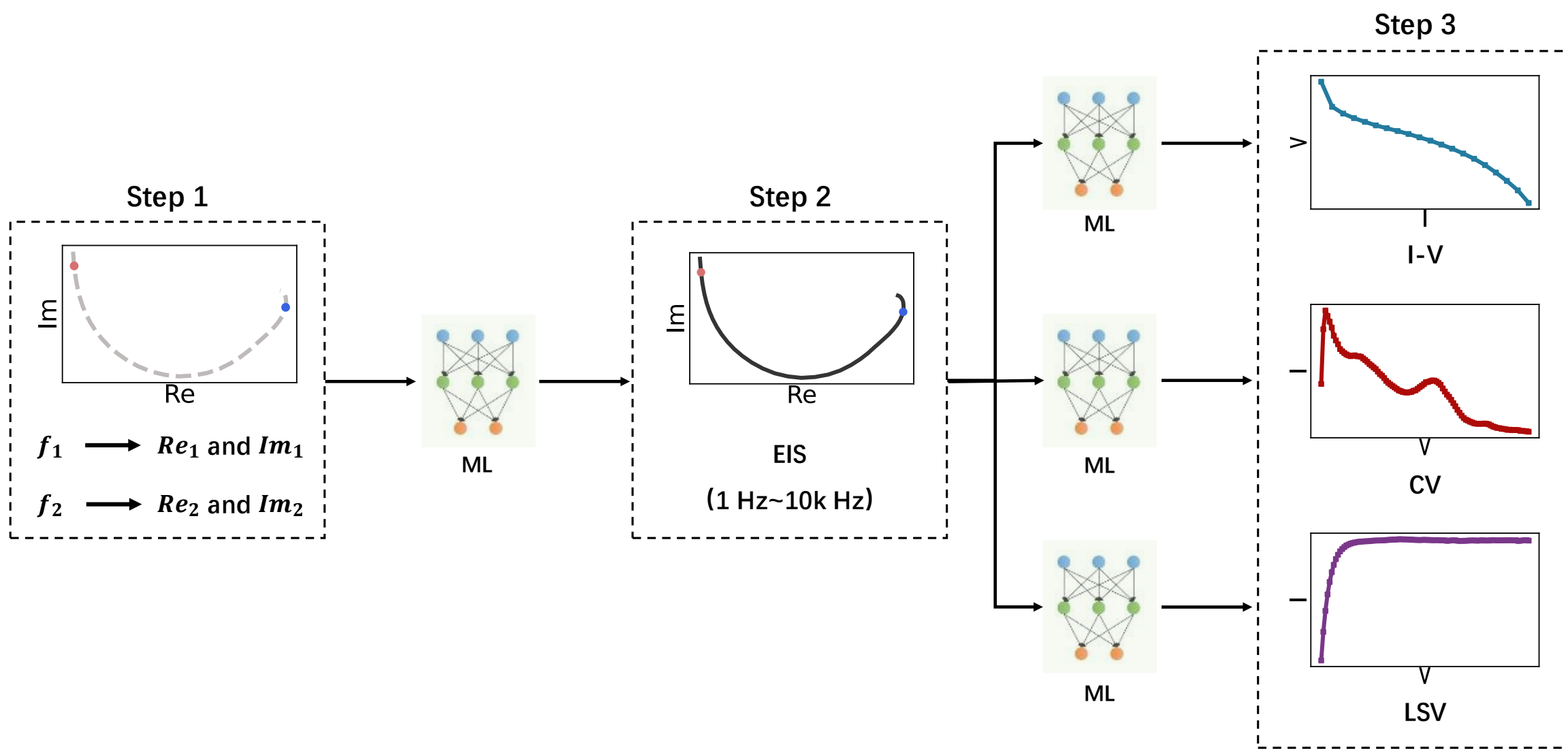

**Figure 1** Steps of the PCDP method.

**Step 1: Impedances measurement.**

The four impedances corresponding to two preset frequencies of the PEMFC is measured under the preset test condition for EIS measurement. The key to this step is to select two preset frequencies, namely a high frequency and a medium frequency. In this paper, we use the machine learning (ML) clustering algorithm to select two preset frequencies from the medium frequency range (1 Hz~100 Hz) and high frequency range (100 Hz~10k Hz) of the real impedance (Re) vs. frequency (Re/f) curve. We name the preset frequency selected from the medium frequency range as $f_1$, and the preset frequency selected from the high frequency range as $f_2$. The two Res corresponding to $f_1$ and $f_2$ are named as $Re_1$ and $Re_2$, respectively. The two imaginary impedances (Ims) corresponding to $f_1$ and $f_2$ are named as $Im_1$ and $Im_2$, respectively. In Dataset 1, the two preset frequencies are selected as $f_1$= 7.9433 Hz and $f_2$ = 7943.3 Hz, as shown in **Figure 2**. For more details of this step, please refer to **Supplementary Note 1**.

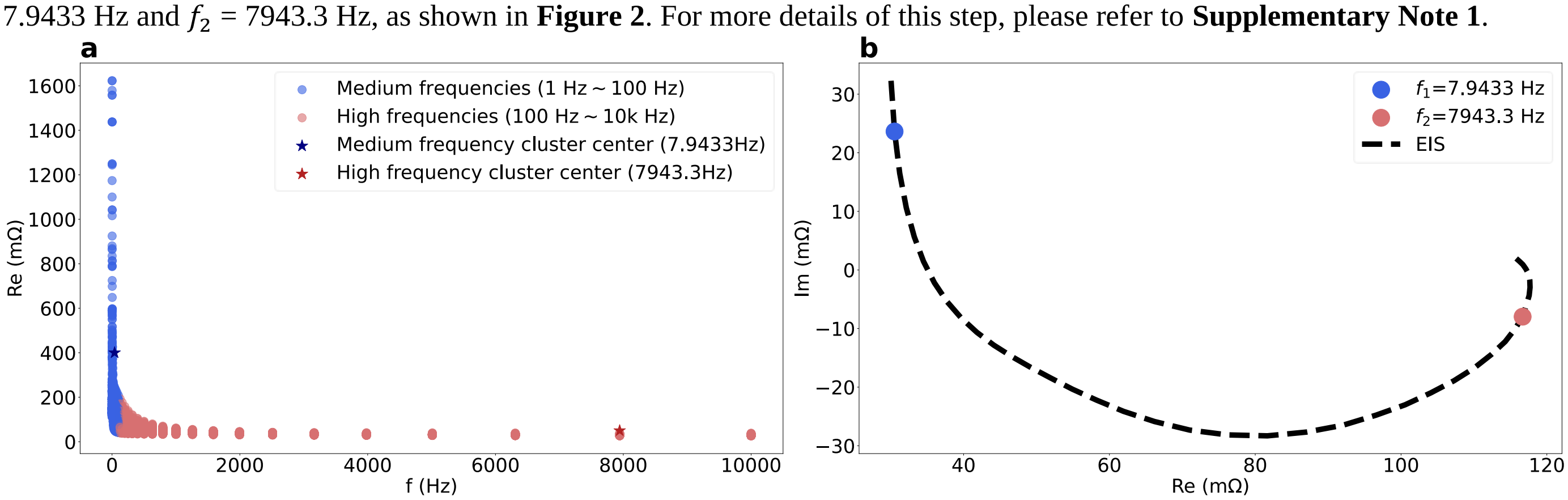

**Figure 2** Test results of Step 1 on the training set of Dataset 1. a. The clustering results in the medium frequency range (1 Hz~100 Hz) and the high frequency range (100 Hz~10k Hz) of Re/f curve. b. The positions of the two preset frequencies $f_1$ and $f_2$ on the EIS of the representative PEMFC (numbered as 1 in the training set of Dataset 1).

**Step 2: EIS prediction.**

Input $Re_1$, $Re_2$, $Im_1$, and $Im_2$ into the EIS prediction model, and output the prediction results of EIS in the medium and high (mid-high) frequency range (1 Hz~10k Hz), as shown in **Figure 3**. The mean absolute error (MEA), root mean square error (RMSE), mean absolute percentage error (MAPE), and coefficient of determination ($R^2$) of the EIS prediction results the on the test set of Dataset 1 are 1.63 $m\Omega$, 2.14, 20.51 %, and 0.97, respectively. Note that in this step, EIS data from 5 test stages, namely 0, 1k, 5k, 10k, and 30k test cycles, are used to train and test the EIS prediction model. For more details of this step, please refer to **Supplementary Note 2**.

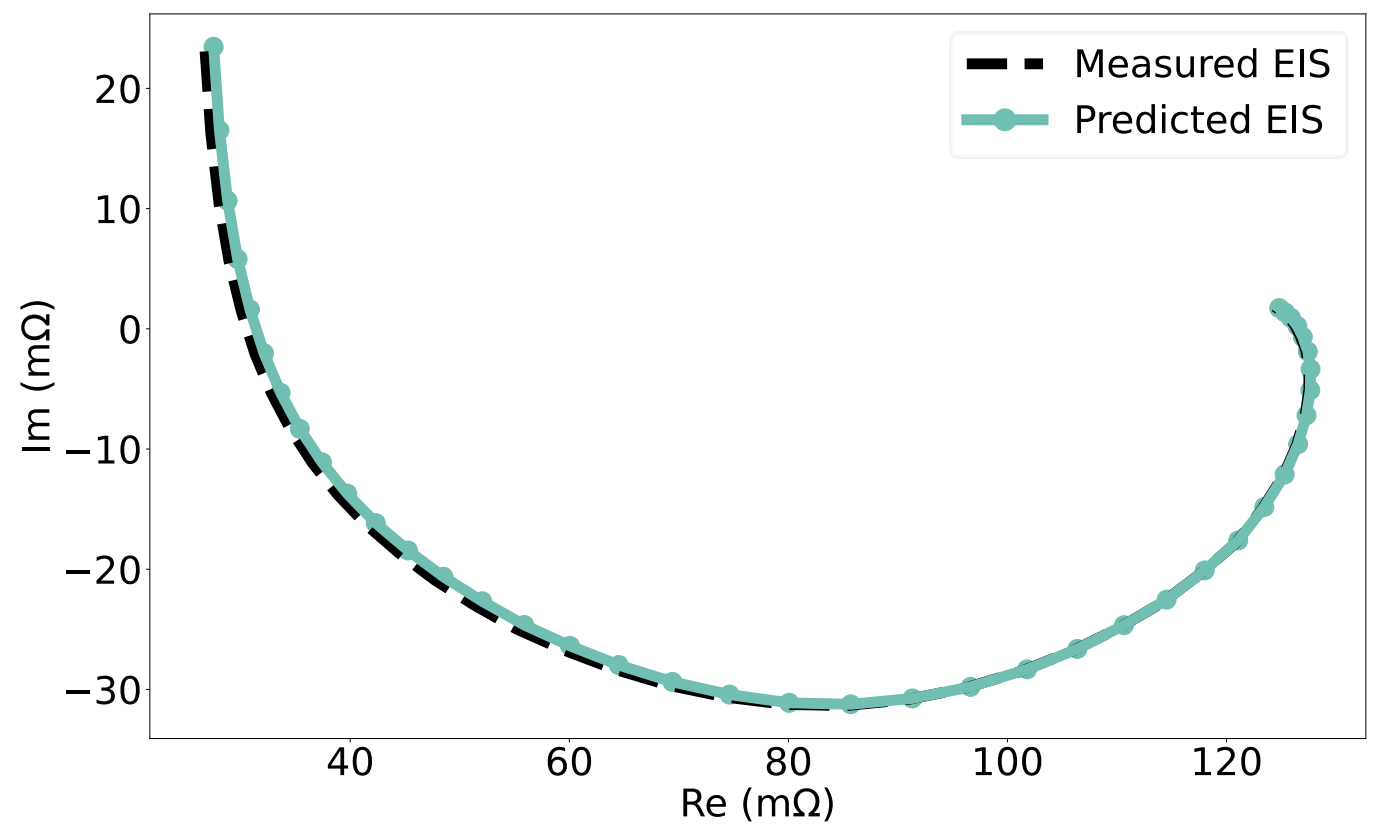

**Figure 3** Test results of Step 2 on the representative cell (numbered as 3 in the test set of Dataset 1). The prediction results of EIS in the mid-high frequency range (1 Hz~10k Hz).

**Step 3: PCD prediction.**

The predicted EIS (1 Hz~10k Hz) is input into the I-V curve prediction model, CV curve prediction model, and LSV curve prediction model, respectively, and the prediction results of I-V curve, CV curve, and LSV curve under different preset test conditions are output, as shown in **Figure 4**. The MEA, RMSE, MAPE, and $R^2$ of the I-V curve prediction results on the test set of Dataset 1 are 0.04V, 0.06, 51.55%, and 0.74, respectively. The MEA, RMSE, MAPE, and $R^2$ of the CV curve prediction results on the test set of Dataset 1 are 2.23 $Acm^{-2}$, 3.19, 14.16%, and 0.50, respectively. The MEA, RMSE, MAPE, and $R^2$ of the LSV curve prediction results on the test set of Dataset 1 are 0.37 $Acm^{-2}$, 0.56, 16.39%, and 0.43, respectively. Note that in this step, I-V curves, CV curves, and LSV curves from 5 test stages, namely 0, 1k, 5k, 10k, and 30k test cycles, are used to train and test the I-V curve prediction model, CV curve prediction model, and LSV curve prediction model, respectively. For more details of this step, please refer to **Supplementary Note 3**.

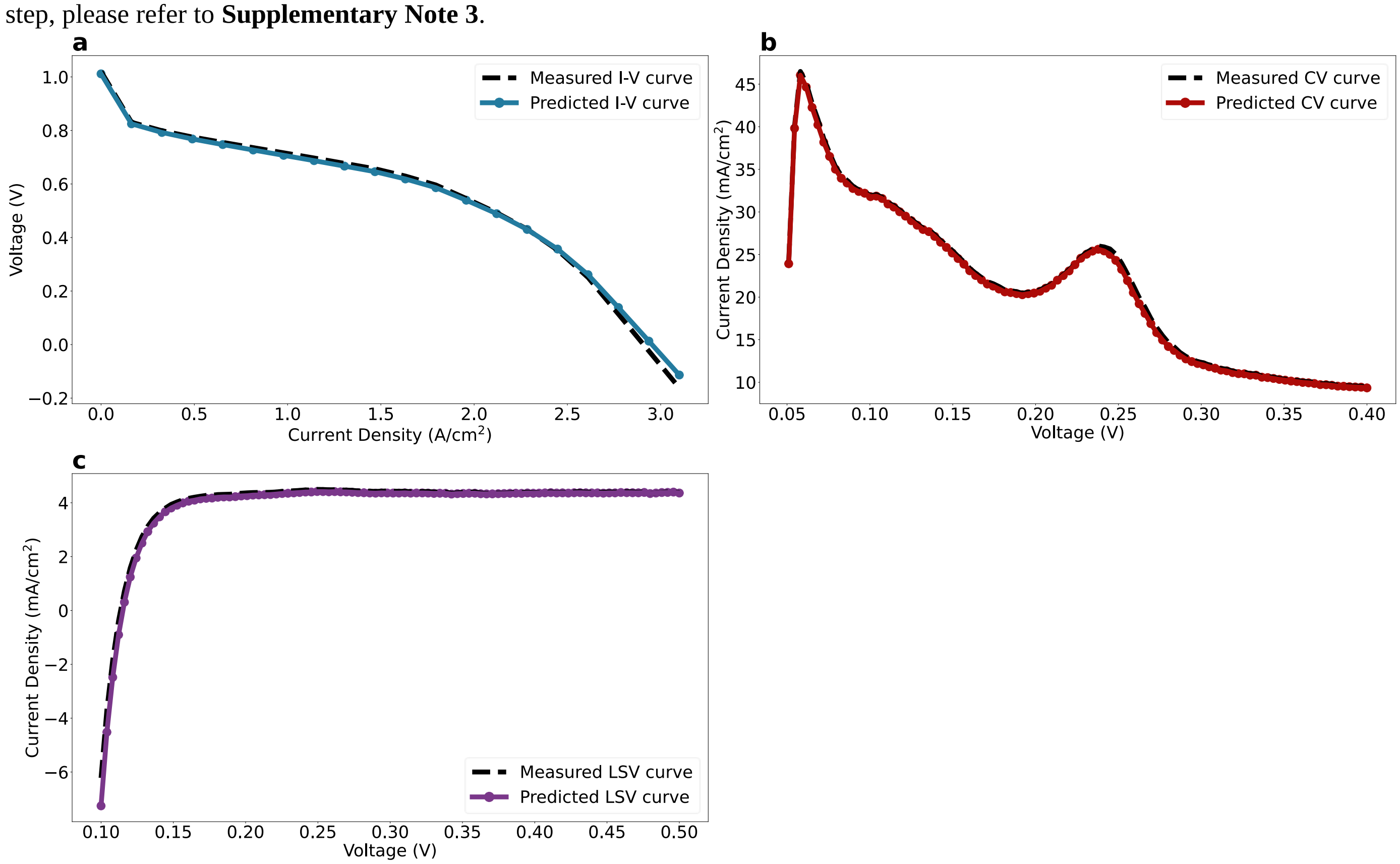

**Figure 4** Test results of Step 3 on the representative cell (numbered as 3 in the test set of Dataset 1). a. The predicted I-V curve and the measured I-V curve. b. The predicted CV curve and the measured CV curve. c. The predicted LSV curve and the measured LSV curve.

To further verify the accuracy of the prediction results of PCD, the different predicted PCD are input into the prediction models of different aging indicators respectively, and the prediction results of different aging indicators are output, as shown in **Figure 5**. The aging indicator prediction models include the $R_{O2,total}$ prediction model, $I_{lim}$ prediction model, $ECSA$ prediction model, and $I_{cross}$ prediction model. Based on the predicted EIS, the MEA, RMSE, MAPE, and $R^2$ of the $R_{O2,total}$ prediction results on the test set of Dataset 1 are 54.61 s/m, 113.97, 21.49%, and 0.69, respectively. Based on the measured EIS, the MEA, RMSE, MAPE, and $R^2$ of the $R_{O2,total}$ prediction results on the test set of Dataset 1 are 47.31 s/m, 105.94, 17.61%, and 0.73, respectively. Based on the predicted I-V curve, the MEA, RMSE, MAPE, and $R^2$ of the $I_{lim}$ prediction results on the test set of Dataset 1 are 0.04 $Acm^{-2}$, 0.06, 21.06%, and 0.79, respectively. Based on the measured I-V curve, the MEA, RMSE, MAPE, and $R^2$ of the $I_{lim}$ prediction results on the test set of Dataset 1 are 0.03 $Acm^{-2}$, 0.06, 19.60%, and 0.80, respectively. Based on the predicted CV curve, the MEA, RMSE, MAPE, and $R^2$ of the $ECSA$ prediction results on the test set of Dataset 1 are 8.30 $cm_{Pt}^2 \cdot cm_{geo}^{-2}$, 14.37, 31.33%, and 0.92, respectively. Based on the measured CV curve, the MEA, RMSE, MAPE, and $R^2$ of the $ECSA$ prediction results on the test set of Dataset 1 are 4.42 $cm_{Pt}^2 \cdot cm_{geo}^{-2}$, 8.59, 16.25%, and 0.97, respectively. Based on the predicted LSV curve, the MEA, RMSE, MAPE, and $R^2$ of the $I_{cross}$ prediction results on the test set of Dataset 1 are 0.18 $Acm^{-2}$, 0.22, 4.55%, and 0.88, respectively. Based on the measured LSV curve, the MEA, RMSE, MAPE, and $R^2$ of the $I_{cross}$ prediction results on the test set of Dataset 1 are 0.12 $Acm^{-2}$, 0.15, 3.21%, and 0.94, respectively. Compared with $R^2$ results of predicted $R_{O2,total}$, $I_{lim}$, and $ECSA$, and $I_{cross}$ based on the actual measured PCD, the $R^2$ results of predicted aging indicators based on the predicted PCD is only reduced by 0.04, 0.01, 0.05, and 0.06, respectively. Note that in this step, since only the measured values of $R_{O2,total}$, $I_{lim}$, and $I_{cross}$ from 2 test stages, namely 0 and 30k test cycles, are provide in Dataset 1, different aging indicators from 2 test stages are used to train and test the $R_{O2,total}$ prediction model, $I_{lim}$ prediction model, $ECSA$ prediction model, and $I_{cross}$ prediction model. For more details, please refer to **Supplementary Note 4**.

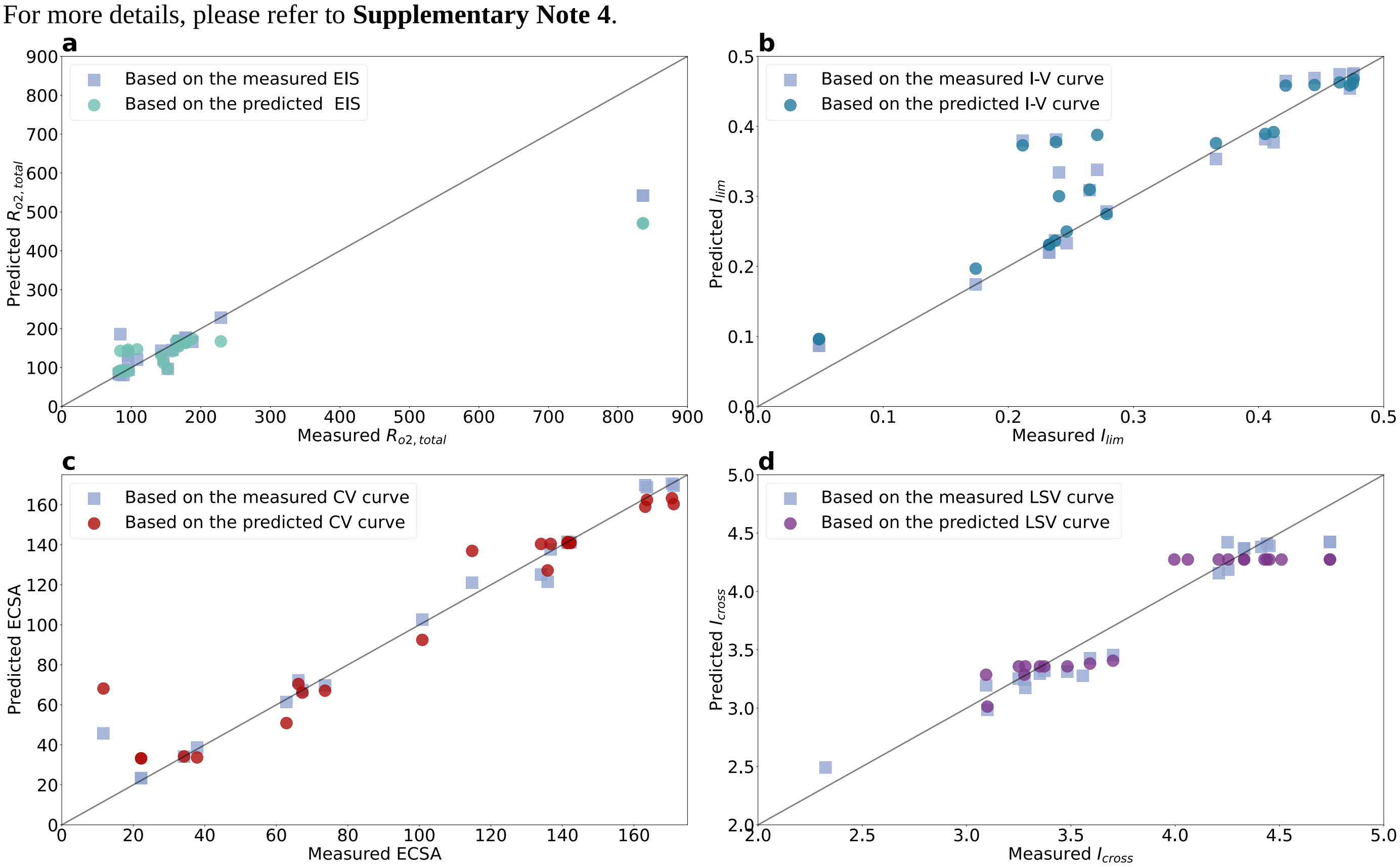

**Figure 5** Prediction results of different PCD on the test set of Dataset 1. a. $R_{O2,total}$ prediction results based on the predicted/measured EIS. b. $I_{lim}$ prediction results based on the predicted/measured I-V curve. c. $ECSA$ prediction results based on the predicted/measured CV curve. d. $I_{cross}$ prediction results based on the predicted/measured LSV curve.

In the PCDP method, all PCD and aging indicator prediction models adopt the random forest (RF) regression model [37]. The hyperparameter settings and training process of RF models are detailed in **Supplementary Note 5**. The evaluation metrics for prediction results are detailed in **Supplementary Note 6**. Test results on Dataset 1 show that the PCDP method can accelerate the offline PCD measurements in the life test of PEMFCs.

## Generalization of PCDP method

In Dataset 2, the two preset frequencies are selected as $f_1$= 6530.0005 Hz and $f_2$ =39.138828 Hz, as shown in **Supplementary Figure 5**. The EIS prediction results on the test set of Dataset 2 are shown in **Supplementary Figure 6**, and the MEA, RMSE, MAPE, and $R^2$ are 0.36 $m\Omega$, 0.28, 8.41 %, and 0.99, respectively. The I-V curve prediction results on the test set of Dataset 2 are shown in **Supplementary Figure 7**, and the MEA, RMSE, MAPE, and $R^2$ are 0.45e-2 V, 0.71e-2, 0.23 %, and 0.98, respectively. Test results on Dataset 2 show that the proposed PCDP method has good generalization performance.

### LP-ALT method

Inspired by [19], the proposed LP-ALT method consists of four steps, namely, PCD collection, difference curve calculation, feature extraction, and life test prediction, is proposed, as shown in **Figure 6**.

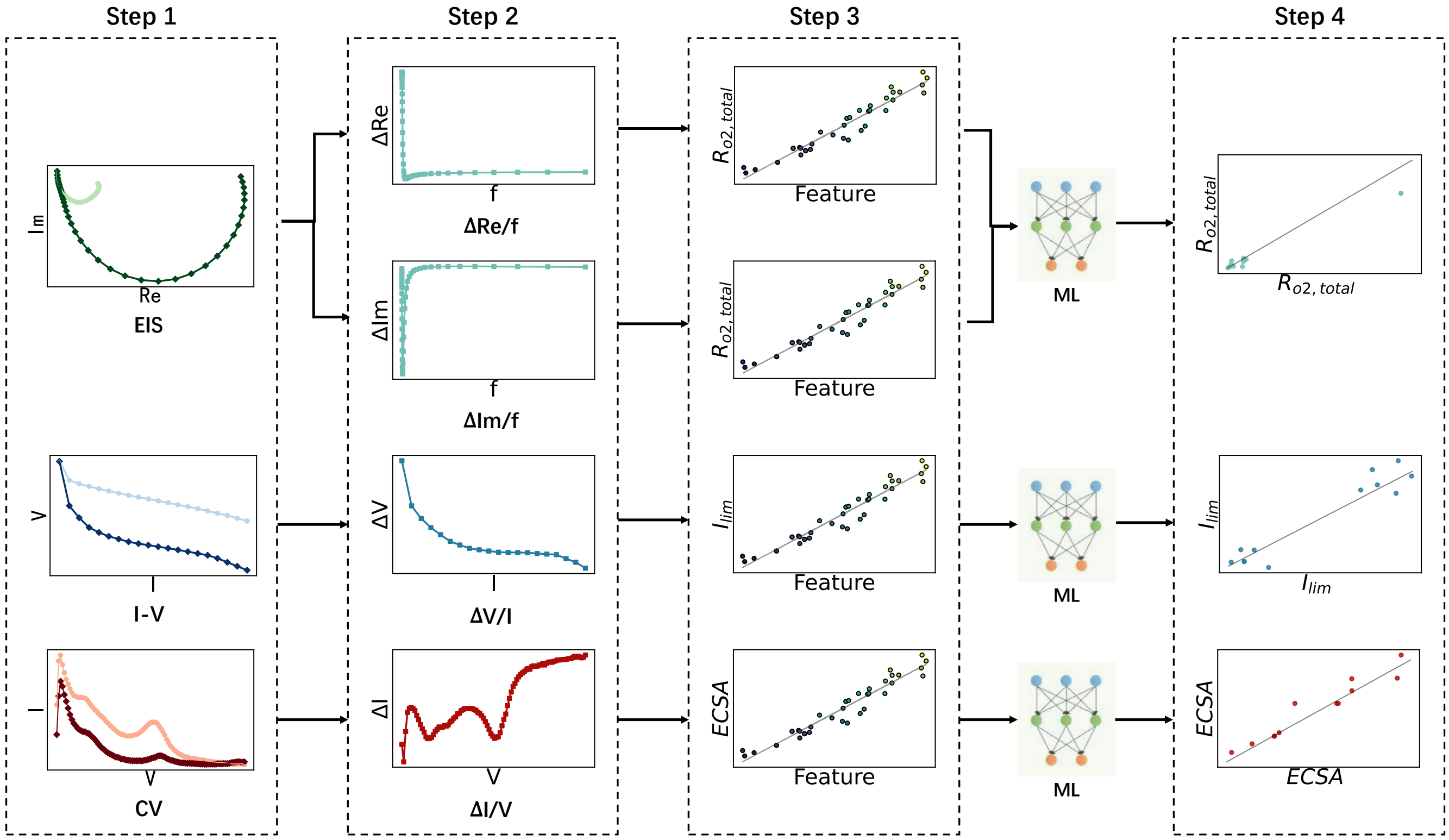

**Figure 6** Steps of the LP-ALT method.

**Step 1: PCD collection.**

Different PCD are collected at two different test stages during the life test of PEMFCs. We named the two different test stages as T1 stage and T2 stage. In Dataset 1, T1 stage is the test time corresponding to the 0th ATC (i.e. 0 s), and T2 stage is the test time corresponding to the 1000th ATC (i.e. 6500 s). The PCD of PEMFCs include I-V curve, EIS, and CV curve. The different PCD collected at two different test stages are shown in **Supplementary Figure 1**.

**Step 2: Difference curve calculation.**

First, the different PCD are standardized. Specifically, the PCD is first fitted using a spline function, and linear interpolation is performed to obtain the PCD after standardization. In Dataset 1, for the I-V curve, the voltage is fitted as a function of the current density using a spline function, and the current density is linearly interpolated. The range of current density values is set to [0mA, 3.1mA], and the linear interpolation interval is set to 0.16mA, as shown in **Supplementary Figure 8 (a).** For the CV curve, the current density is fitted as a function of voltage using a spline function, and the voltage is linearly interpolated. The range of voltage values is set to [0.051V, 0.4V], and the linear interpolation interval is set to 0.003V, as shown in **Supplementary Figure 8 (b)**. For EIS, since the frequencies of EIS data have been set to the same during collection, different EIS data can be directly subtracted without standardization.

Then, the PCD collected in the T2 stage are subtracted from the corresponding PCD collected in the T1 stage to obtain the difference curves. The difference curves corresponding to the I-V curve and the CV curve are the voltage difference vs. current density (ΔV/I) curve and the current density difference vs. voltage (ΔI/V) curve, respectively. The difference curves corresponding to the EIS are the Re difference (ΔRe/f) curve and the Im difference (ΔIm/f) curve. Specifically, the ΔV/I curve is calculated by subtracting the standardized I-V curve of the T1 stage from the standardized I-V curve of the T2 stage, as shown in **Supplementary Figure 9**. The ΔI/V curve is calculated by subtracting the standardized CV curve of the T2 stage from the standardized CV curve of the T1 stage, as shown in **Supplementary Figure 10**. The ΔRe/f curve is calculated by subtracting the Re/f curve of the T1 stage from the Re/f curve of the T2 stage, and the ΔIm/f curve is calculated by subtracting the Im/f curve of the T1 stage from the Im/f curve of the T2 stage, as shown in **Supplementary Figures 11**.

**Step 3: Feature extraction.**

Features related to the aging indicators corresponding to the predicted target stage (named as T3 stage) are extracted from difference curves, such as instantaneous features, statistical features, and model parameter features. In Dataset 1, the T3 stage is the test time corresponding to the 30,000th ATC (i.e., 125 h). The features extracted and selected from the ΔV/I curve, ΔRe/f curve,

ΔIm/f curve, and ΔI/V curve are all the two-point sure independence screening and sparsifying operator (TP-SISSO) feature, which are respectively related to the values of the aging indicators $I_{lim}$, $R_{o2,total}$, and $ECSA$ in the T3 stage.

Specifically, the methods for extracting the TP-SISSO feature from the ΔV/I curve is as follows: (1) Subtract the voltages corresponding to any two current densities on the ΔV/I curve, and take the absolute value as a candidate two-point feature; (2) Traverse all combinations of two current densities on the ΔV/I curve, and a total of $(n^2$-$n)$/2 candidate two-point features can be obtained, where $n$ is the number of current densities on the ΔV/I curve and $n$=20 in Dataset 1; (3) The operators of the SISSO model are defined as the addition and subtraction. All candidate two-point features and the corresponding aging indicator $I_{lim}$ in the T3 stage are integrated into the input data format of the SISSO model. The SISSO model automatically calculates and outputs the TP-SISSO feature calculation formula corresponding to the $I_{lim}$ prediction in the T3 stage. In Dataset 1, the relationship between the TP-SISSO feature extracted from the ΔV/I curve and the change in $I_{lim}$ from the T1 stage to T3 stage is shown in **Supplementary Figure 12**.

The methods for extracting BTPF and TP-SISSO feature from the ΔRe/f curve is as follows: (1) Subtract the Res corresponding to any two frequencies on the ΔRe/f curve, and take the absolute value as a candidate two-point feature; (2) Traverse all combinations of two frequencies on the ΔRe/f curve, and a total of $(n^2$-$n)$/2 candidate two-point features can be obtained, where $n$ is the number of frequencies on the ΔRe/f curve and $n$=51 in Dataset 1; (3) The operators of the SISSO model are defined as the addition and subtraction. All candidate two-point features and the corresponding aging indicator $R_{o2,total}$ in the T3 stage are integrated into the input data format of the SISSO model. The SISSO model automatically calculates and outputs the TP-SISSO feature calculation formula corresponding to the $R_{o2,total}$ prediction in the T3 stage. In Dataset 1, the relationship between the TP-SISSO feature extracted from the ΔRe/f curve and the change in $R_{o2,total}$ from the T1 stage to T3 stage is shown in **Supplementary Figure 13(a)**.

The methods for extracting BTPF and TP-SISSO feature from the ΔIm/f curve is the same as that of the ΔRe/f curve. In Dataset 1, the relationship between the TP-SISSO feature extracted from the ΔIm/f curve and the change in $R_{o2,total}$ from the T1 stage to T3 stage is shown in **Supplementary Figure 13(b)**.

The methods for extracting BTPF and TP-SISSO feature from the ΔI/V curve is as follows: (1) Subtract the current densities corresponding to any two voltages on the ΔI/V curve, and take the absolute value as a candidate two-point feature; (2) Traverse all combinations of two voltages on the ΔI/V curve, and a total of $(n^2$-$n)$/2 candidate two-point features can be obtained, where $n$ is the number of voltages on the ΔI/V curve and $n$=100 in Dataset 1; (3) The operators of the SISSO model are defined as the addition and subtraction. All candidate two-point features and the corresponding aging indicator $ECSA$ in the T3 stage are integrated into the input data format of the SISSO model. The SISSO model automatically calculates and outputs the TP-SISSO feature calculation formula corresponding to the $ECSA$ prediction in the T3 stage. In Dataset 1, the relationship between the TP-SISSO feature extracted from the ΔI/V curve and the change in $ECSA$ from the T1 stage to T3 stage is shown in **Supplementary Figure 14**.

For more details about the TP-SISSO feature extraction method, please refer to [38, 39] and **Supplementary Note 7**.

**Step 4: Life test prediction.**

The extracted features are input into the life prediction model, and the model outputs the prediction results of different aging indicators in the T3 stage. Specifically, in Dataset 1, the life prediction model includes the $I_{lim}$ prediction model, the $R_{o2,total}$ prediction model, and the $ECSA$ prediction model. The inputs of the $I_{lim}$ prediction model is the TP-SISSO feature extracted from the ΔV/I curve, and the output is the predicted value of $I_{lim}$ in the T3 stage, as shown in **Figure 7(a)**. The MEA, RMSE, MAPE, and $R^2$ of the $I_{lim}$ prediction results on the test set of Dataset 1 are 0.020 A/$cm^2$, 0.028, 7.98%, and 0.94, respectively. The input of the $R_{o2,total}$ prediction model are the TP-SISSO features extracted from the ΔRe/f and ΔIm/f curves, and the output is the predicted value of $R_{o2,total}$ in the T3 stage, as shown in **Figure 7(b)**. The MEA, RMSE, MAPE, and $R^2$ of the $R_{o2,total}$ prediction results on the test set of Dataset 1 are 28.83 s/m, 55.42, 10.53%, and 0.93, respectively. The input of the $ECSA$ prediction model is the TP-SISSO feature extracted from the ΔI/V curve, and the output is the predicted value of $ECSA$ in the T3 stage, as shown in **Figure 7(c)**. The MEA, RMSE, MAPE, and $R^2$ of the $ECSA$ prediction results on the test set of Dataset 1 are 8.99 $cm_{Pt}^2 \cdot cm_{geo}^{-2}$, 12.12, 18.53%, and 0.89, respectively. For more details of this step, please refer to **Supplementary Note 8**.

As shown in **Figure 7**, the proposed LP-ALT method can accelerate the life test of PEMFCs through life prediction. Compared with the complete life test of 30,000 ATCs, the LP-ALT method only requires 1,000 ATCs, with an acceleration ratio of 30 times, and can ensure that the minimum $R^2$ of the prediction results for different aging indicators are not less than 0.9. Note that the 30 times acceleration ratio is limited by the data provided by Dataset 1. We speculate that the proposed LP-ALT method can achieve a greater acceleration ratio than 30 times in the life test of PEMFCs.

In the LP-ALT method, all life prediction models adopt the RF regression model [36]. The hyperparameter settings and training process of RF models are detailed in **Supplementary Note 5**. The evaluation metrics for prediction results are detailed in **Supplementary Note 6**.

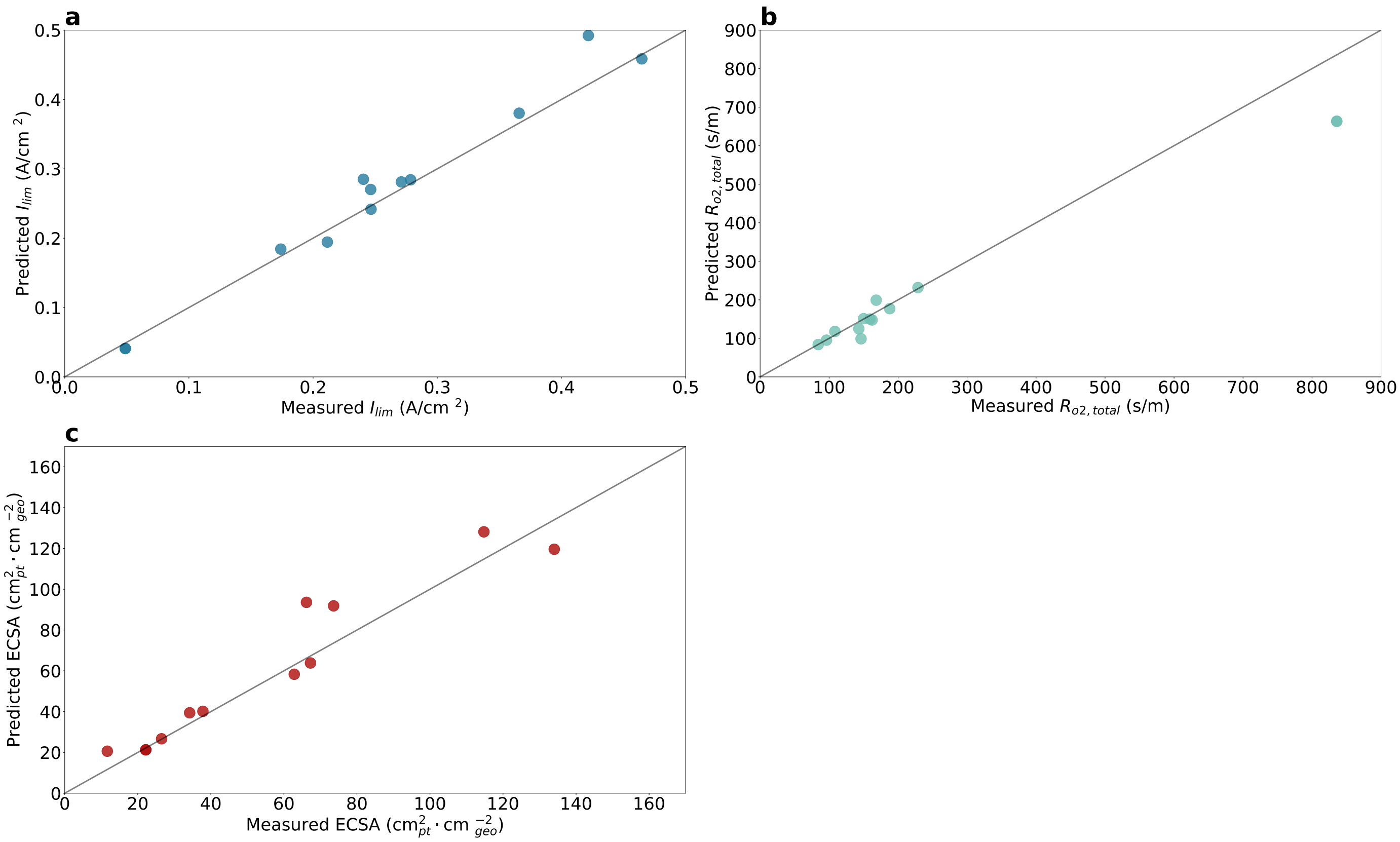

**Figure 7** Prediction results of different aging indicators in the T3 stage on the test set of Dataset 1. a. Prediction results of $I_{lim}$. b. Prediction results of $R_{o2,total}$. c. Prediction results of $ECSA$.

## Generalization of LP-ALT method

The test results and analysis of the LP-ALT method on Dataset 3 are as follows. In Dataset 3, T1 stage is the 0 h, T2 stage is the 125 h, and T3 stage is the 5105.5 h. The PCD of the capacitor is EIS. The aging indicator of the capacitor is the remaining capacitance ($C_{rem}$). EIS data collected at different test stages are shown in **Supplementary Figure 3**. As shown in **Supplementary Figures 16**, the ΔRe/f curve is calculated by subtracting the Re/f curve of the T1 stage from the Re/f curve of the T2 stage. The ΔIm/f curve is calculated by subtracting the Im/f curve of the T1 stage from the Im/f curve of the T2 stage.

The features extracted and selected from the ΔRe/f curve and the ΔIm/f curve are TP-SISSO features, which are related to the value of $C_{rem}$ in the T3 stage. The relationship between the extracted TP-SISSO feature from the ΔRe/f curve and the change in $C_{rem}$ from the T1 stage to T3 stage is shown in **Supplementary Figure 17(a)**. The relationship between the extracted TP-SISSO feature from the ΔIm/f curve and the change in $C_{rem}$ from the T1 stage to T3 stage is shown in **Supplementary Figure 17(b)**.

The life prediction model is the $C_{rem}$ prediction model. The inputs of the $C_{rem}$ prediction model are two TP-SISSO features extracted from the ΔRe/f curve and the ΔIm/f curve, and the output is the predicted value of $C_{rem}$ in the T3 stage, as shown in **Figure 8**. The MEA, RMSE, MAPE, and $R^2$ of the $C_{rem}$ prediction results are 38.72 $\mu F$, 49.51, 2.45%, and 0.74, respectively.

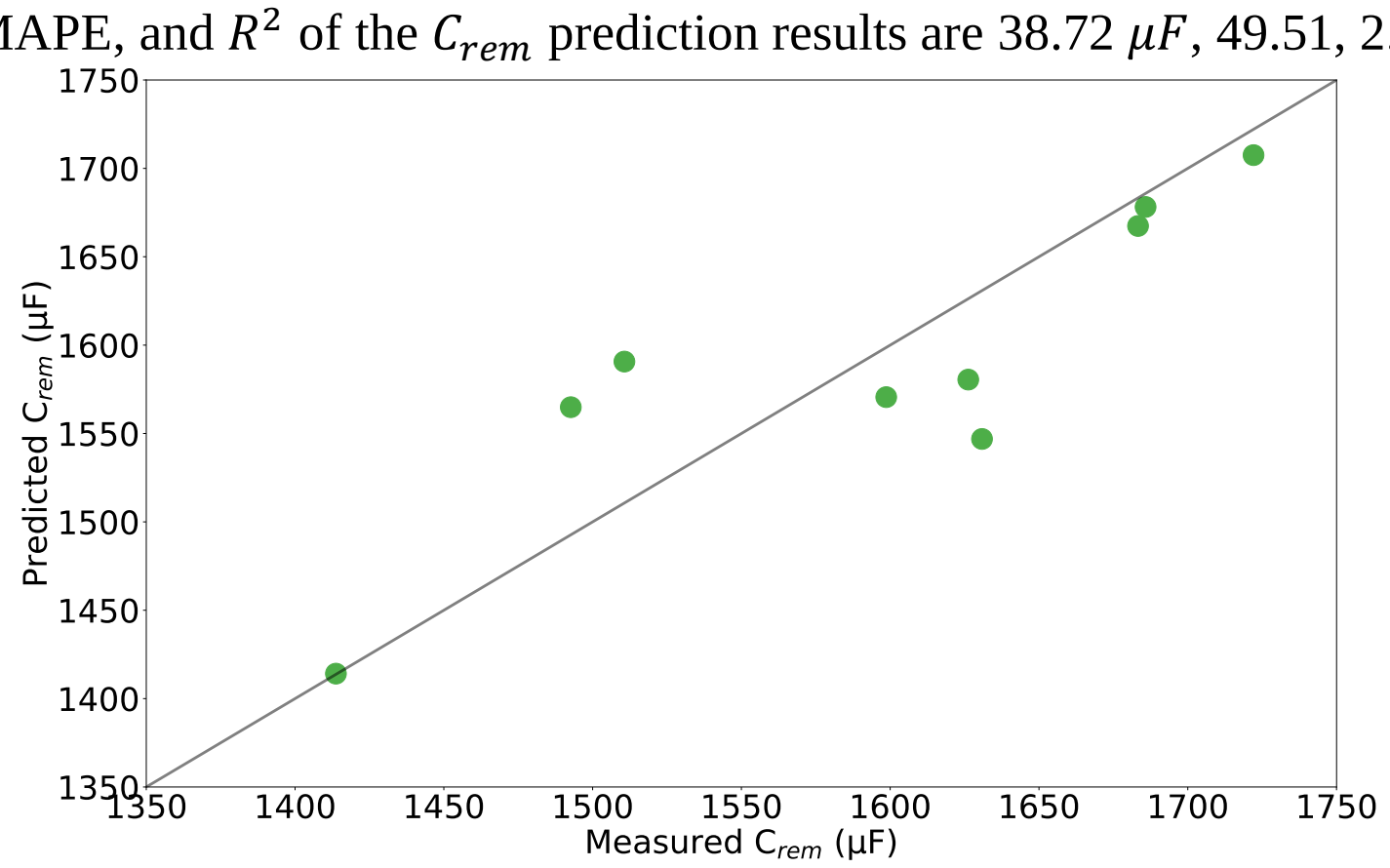

**Figure 8** Prediction results of $C_{rem}$ value in the T3 stage on the test set of Dataset 3.

As shown in **Figure 8**, the LP-ALT method proposed in this paper can accelerate the life test of capacitors through life prediction. Compared with the full life test of 5105.5 h, the LP-ALT method only takes 125 h, with an acceleration ratio of more than 40 times, and can ensure that the MAPE of the prediction result for $C_{rem}$ is 2.45%. Test results on Dataset 3 show that the LP-ALT method has good generalization performance.

## Conclusions

To reduce the economic, time, and labor costs of PEMFC life testing, this paper first proposed the PCDP method. By only collecting the four impedances corresponding to a high frequency and a medium frequency, the EIS, I-V curve, CV curve, and LSV curve can be accurately predicted, which can greatly reduce the time and difficulty of regular offline PCD measurements during PEMFC life testing. Moreover, based on the predicted PCD, accurate predictions of different PEMFC aging indicators were achieved. Then, this paper proposed the LP-ALT method, which can greatly shorten the time of PEMFC life testing through early life prediction. Test results on the same life test dataset containing 42 PEMFCs show that the LP-ALT method can accurately predict the aging indicators at the 30,000th ATC by only collecting PCD at the 0th and 1000th ATC, and the acceleration ratio is 30 times. Furthermore, the generalization of the PCDP and LP-ALT methods is verified using an open-source life test dataset of a PEMWE cell and an open-source life test dataset containing 24 capacitors. Combining the different PCD predicted by the PCDP method and the life prediction of the LP-ALT method, the diagnosis and prognosis of PEMFCs and their components can be realized. This work prioritizes method exploration rather than maximizing prediction accuracy.

## Data availability

All datasets used in this study are publicly accessible and include Dataset 1 (https://doi.org/10.6084/m9.figshare.25450177), D ataset 2 (https://data.uni-hannover.de/zh_Hans_CN/dataset/pemwe-single-cell-dataset-under-ast-protocol), and Dataset 3 (https://phm-datasets.s3.amazonaws.com/NASA/12.+Capacitor+Electrical+Stress.zip).

Source data of this paper will be available after publication.

## Code availability

The code needed to replicate the results and figures in this paper will be available after publication.

## Acknowledgements

This work was funded by the Natural Science Foundation of Zhejiang Province (Grant No. LQ23E050013) and National Natural Science Foundation of China (Grant No. U2141209).

## Author contributions

Y. Zhao: Conceptualization, software, methodology, data curation and analysis, visualization, writing-review and editing.
H. Liu: Conceptualization, resources, software, methodology, data curation and analysis, visualization, writing-review and editing, project administration, supervision, funding acquisition.
Z. Deng: Data curation and analysis, visualization, writing-review and editing.
H. Jiang: Data curation and analysis, visualization, writing-review and editing.
Z. Ling: Data curation and analysis, visualization, writing-review and editing.
Z. Liu: Data curation and analysis, visualization, writing-review and editing.
X. Wang: Writing-review and editing.
T. Li: Writing-review and editing.
X. Ouyang: Writing-review and editing, project administration, funding acquisition.

## Competing interests

The authors declare no competing interests.

Supplementary Information

# Machine learning accelerates fuel cell life testing

Yanbin Zhao[†1], Hao Liu[†*1], Zhihua Deng[2], Haoyi Jiang[1], Zhenfei Ling[1], Zhiyang Liu[3], Xingkai Wang[4], Tong Li[1], Xiaoping Ouyang[1]

[1] State Key Laboratory of Fluid Power and Mechatronic Systems, School of Mechanical Engineering, Zhejiang University, Hangzhou, China.
[2] Energy Research Institute at NTU (ERI@N), Nanyang Technological University, Singapore.
[3] School of Control Science and Engineering, Zhejiang University, Hangzhou, China.
[4] School of Materials Science and Engineering, Tianjin University, Tianjin, China.
[†] These authors contributed equally to this work: Yanbin Zhao, Hao Liu.
[*] Corresponding author's email: haoliu7850052@zju.edu.cn (Hao Liu).

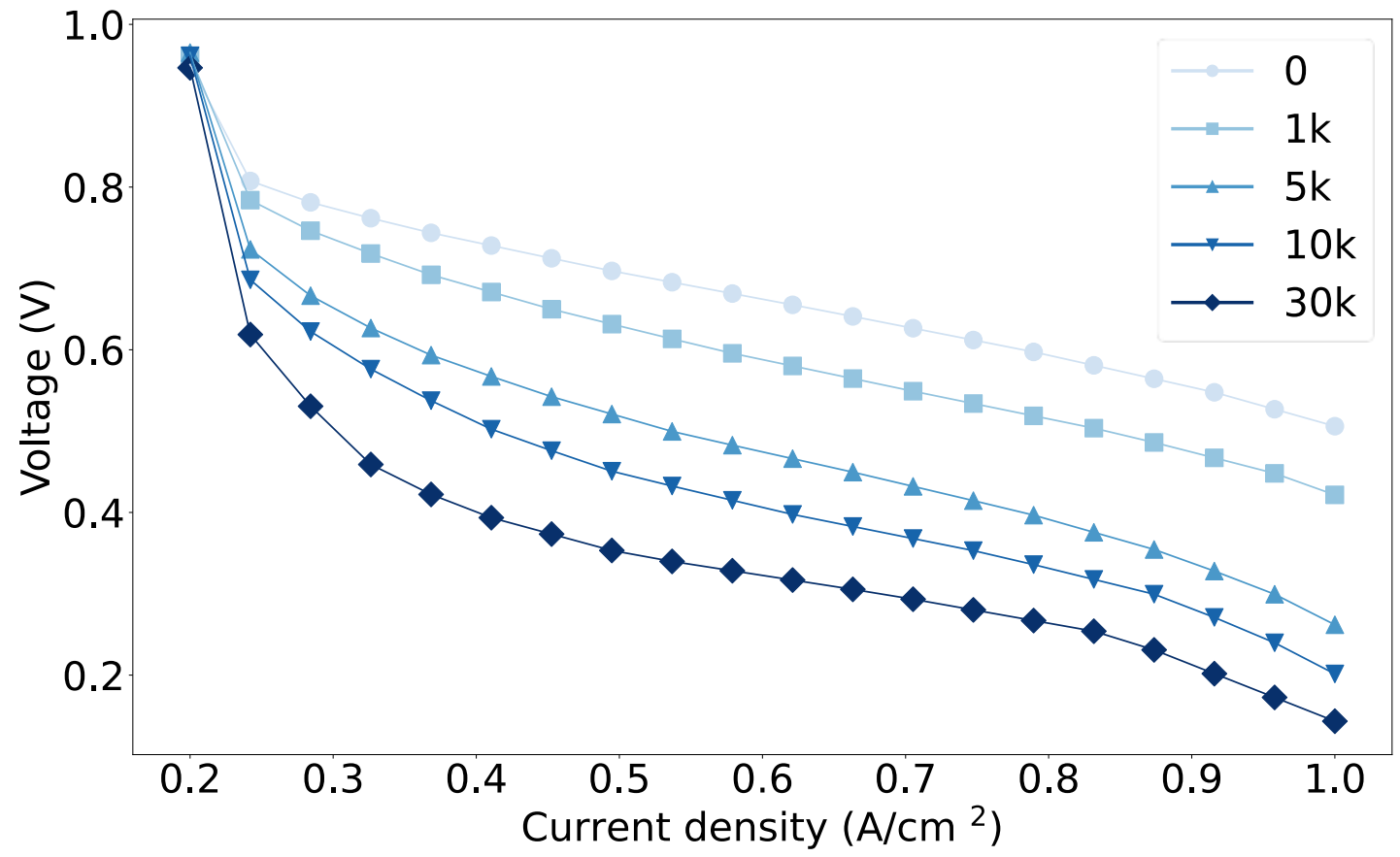

(a) I-V curves

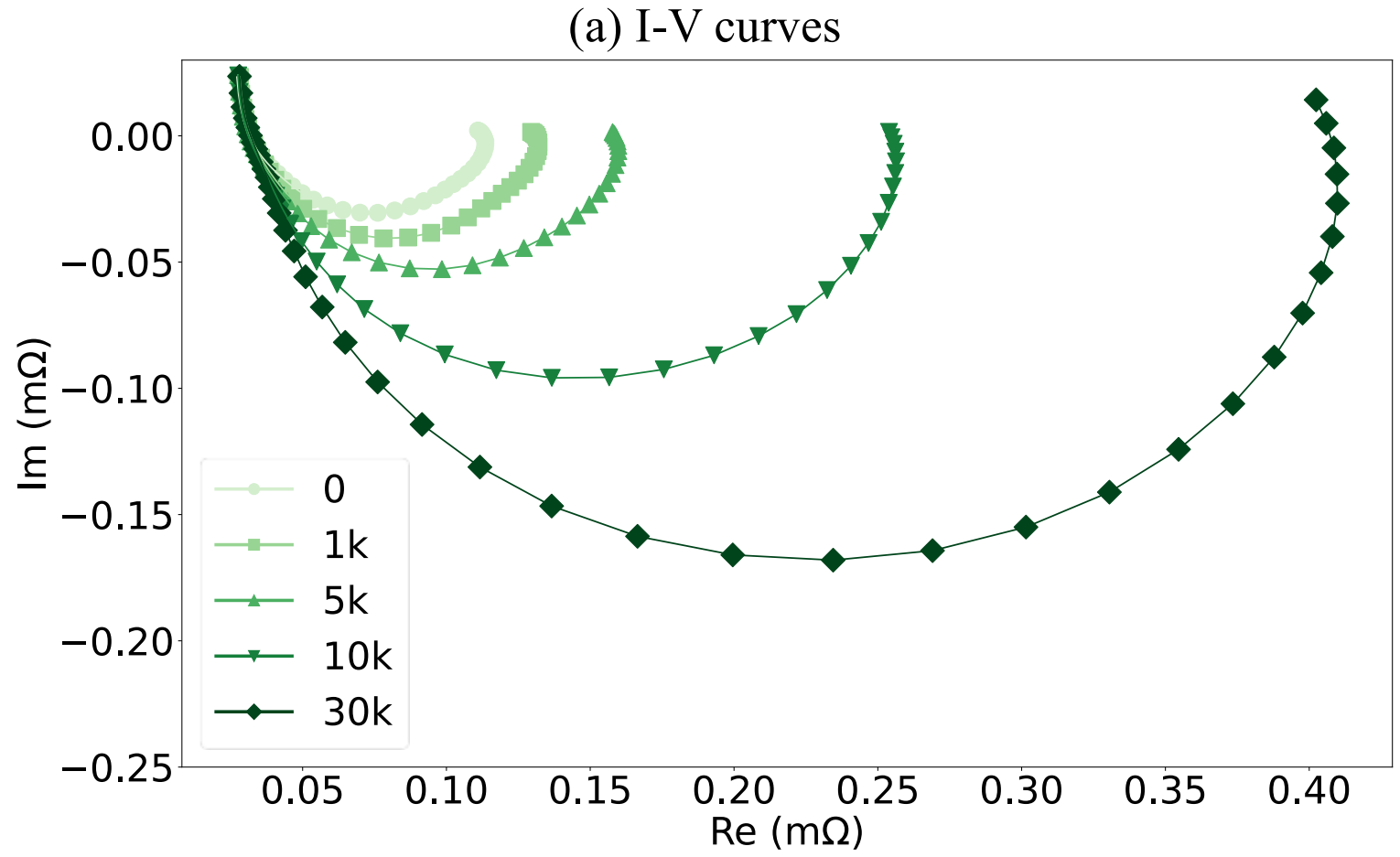

(b) EIS

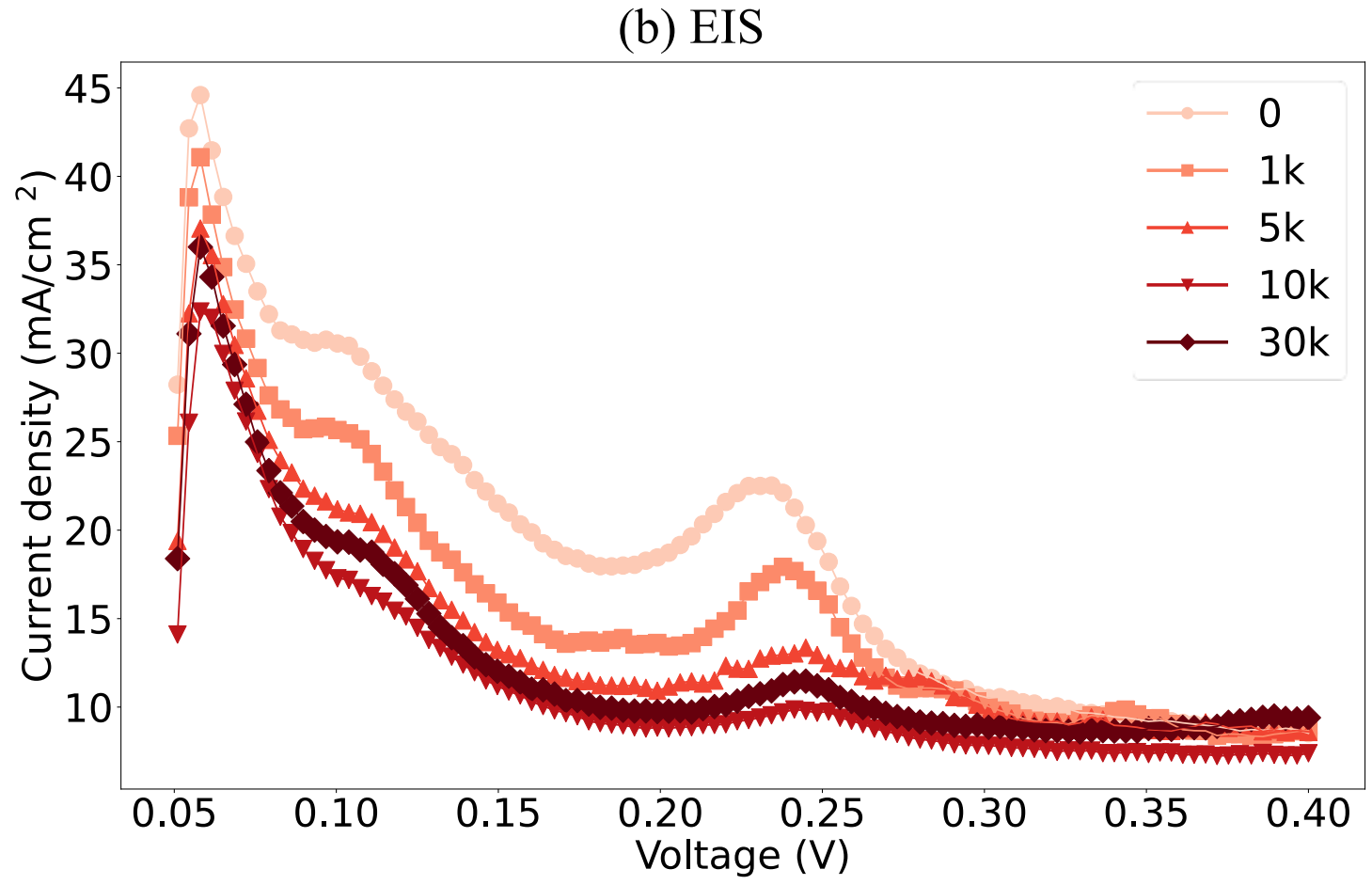

(c) CV curves

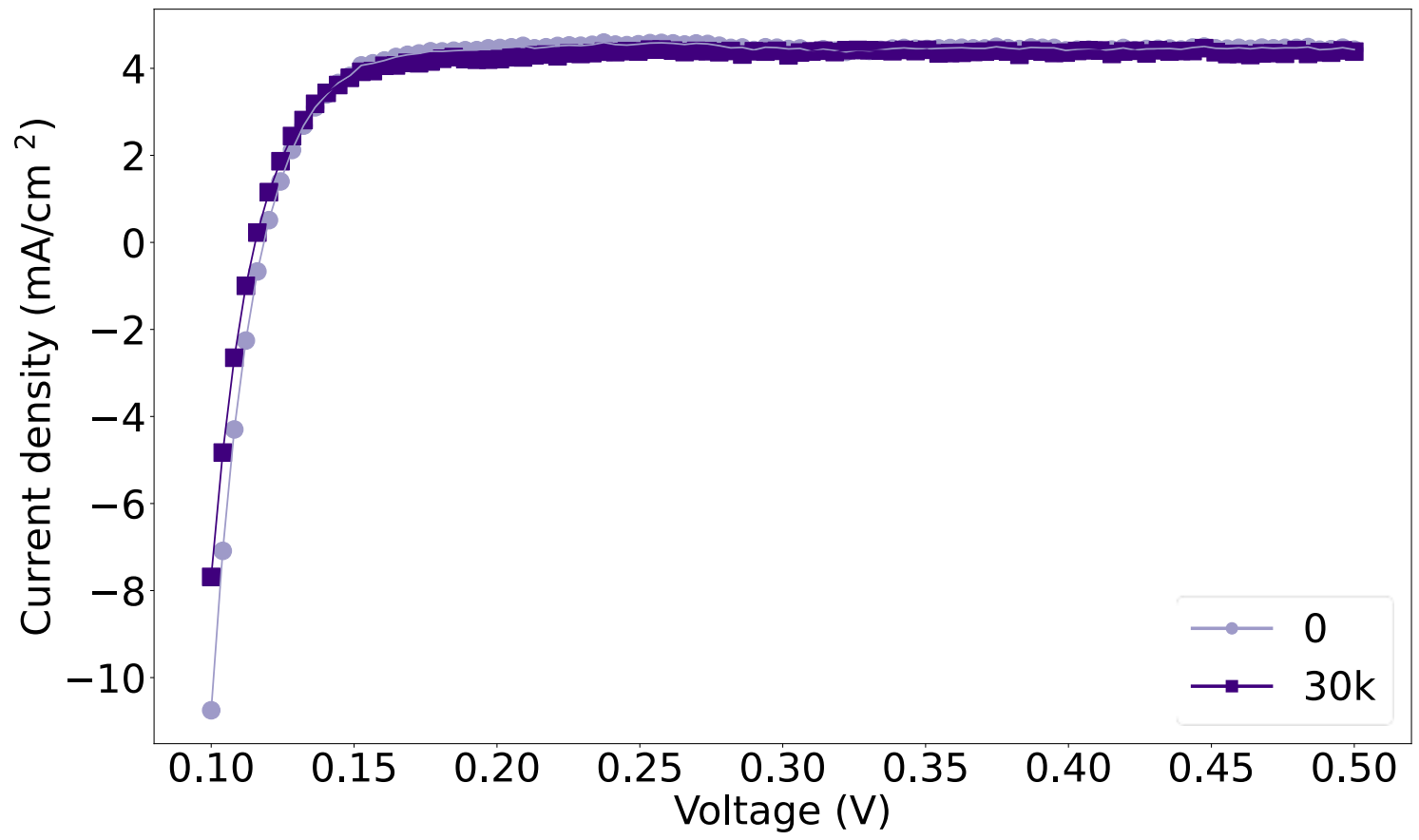

(d) LSV curves

**Supplementary Figure 1** Different PCD of the representative PEMFC (numbered as 3 in the test set of Dataset 1) measured at different test stages.

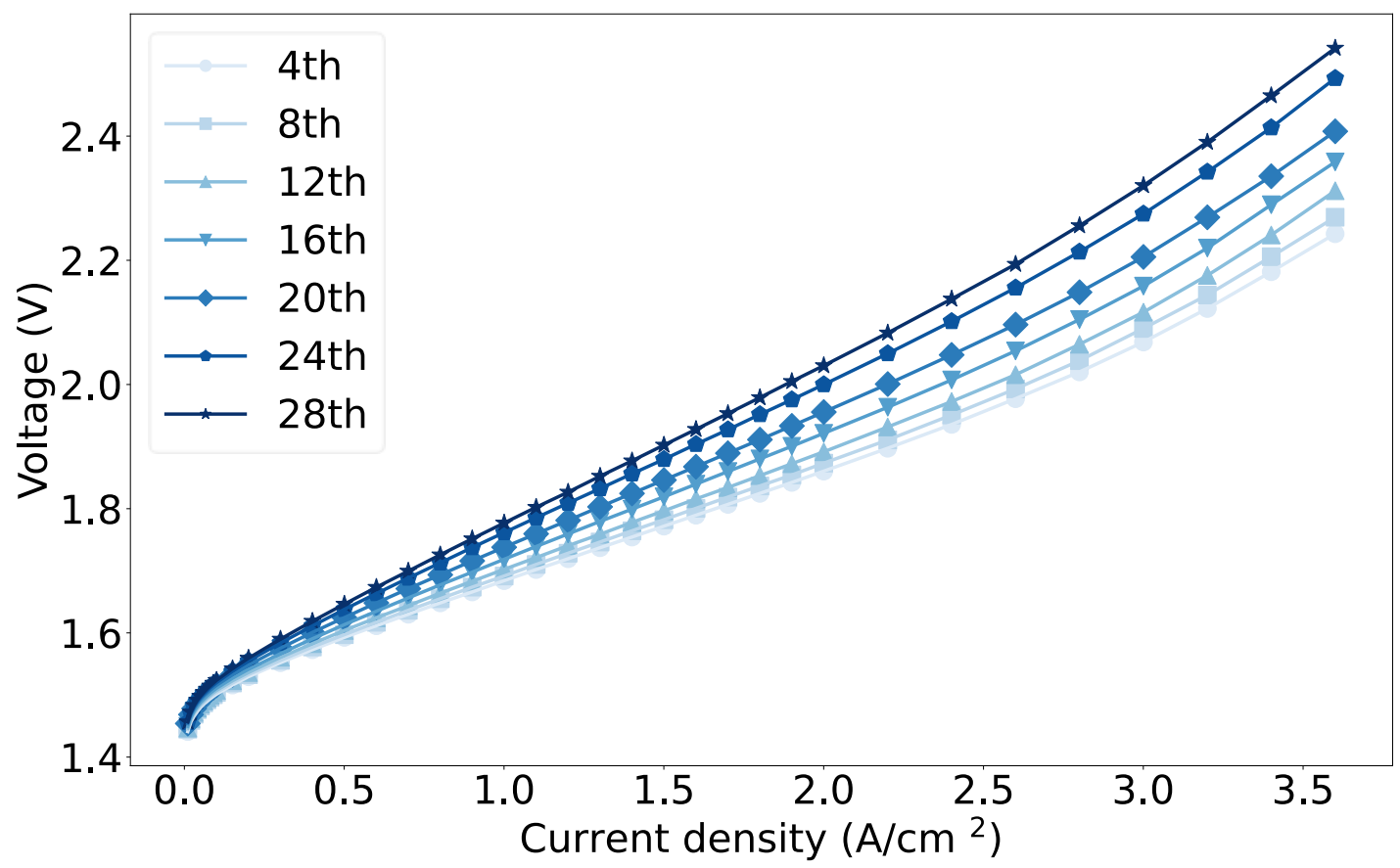

(a) I-V curves

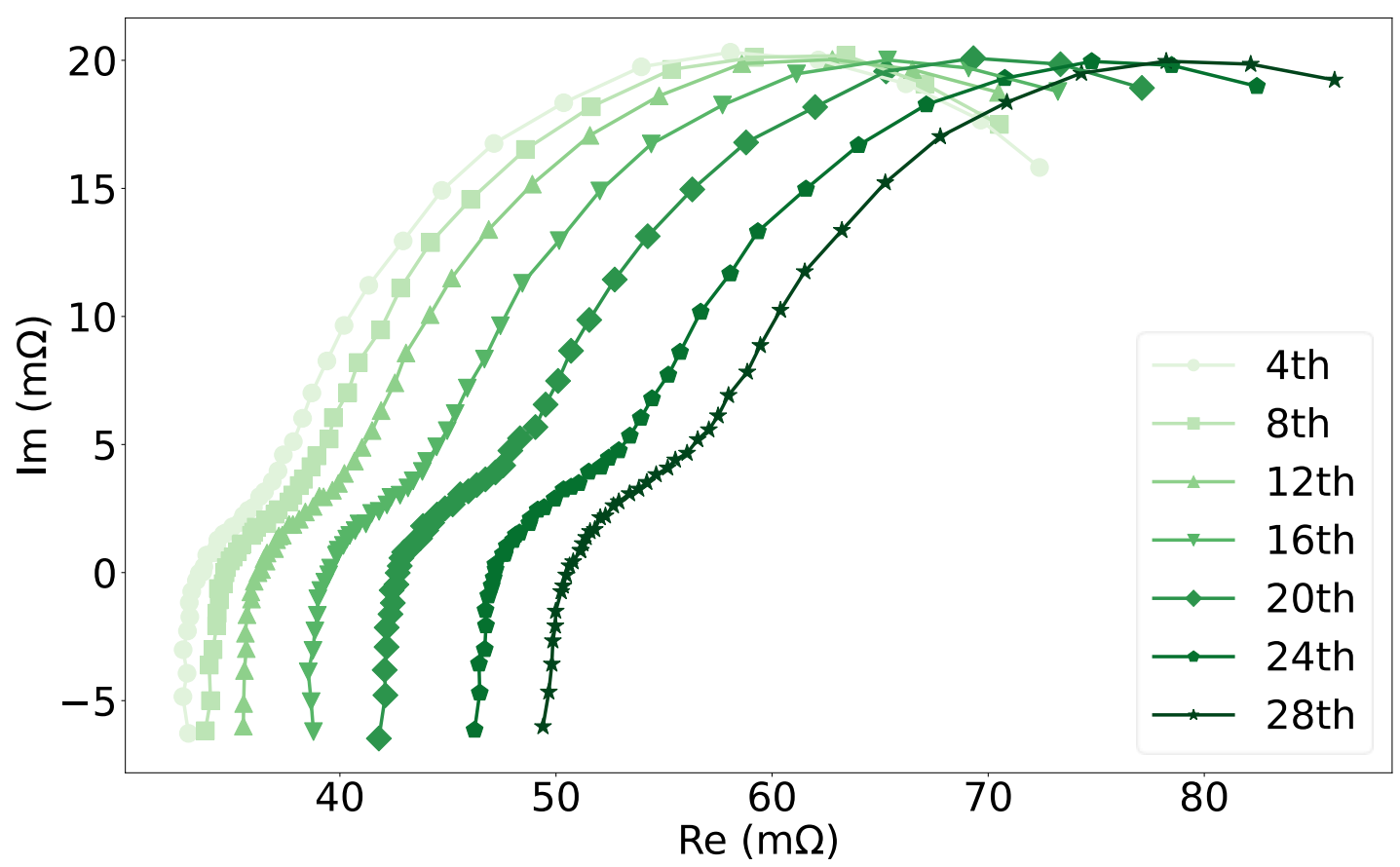

(b) EIS

**Supplementary Figure 2** Different PCD of the PEMWE cell measured at different test stages in the test set of Dataset 2.

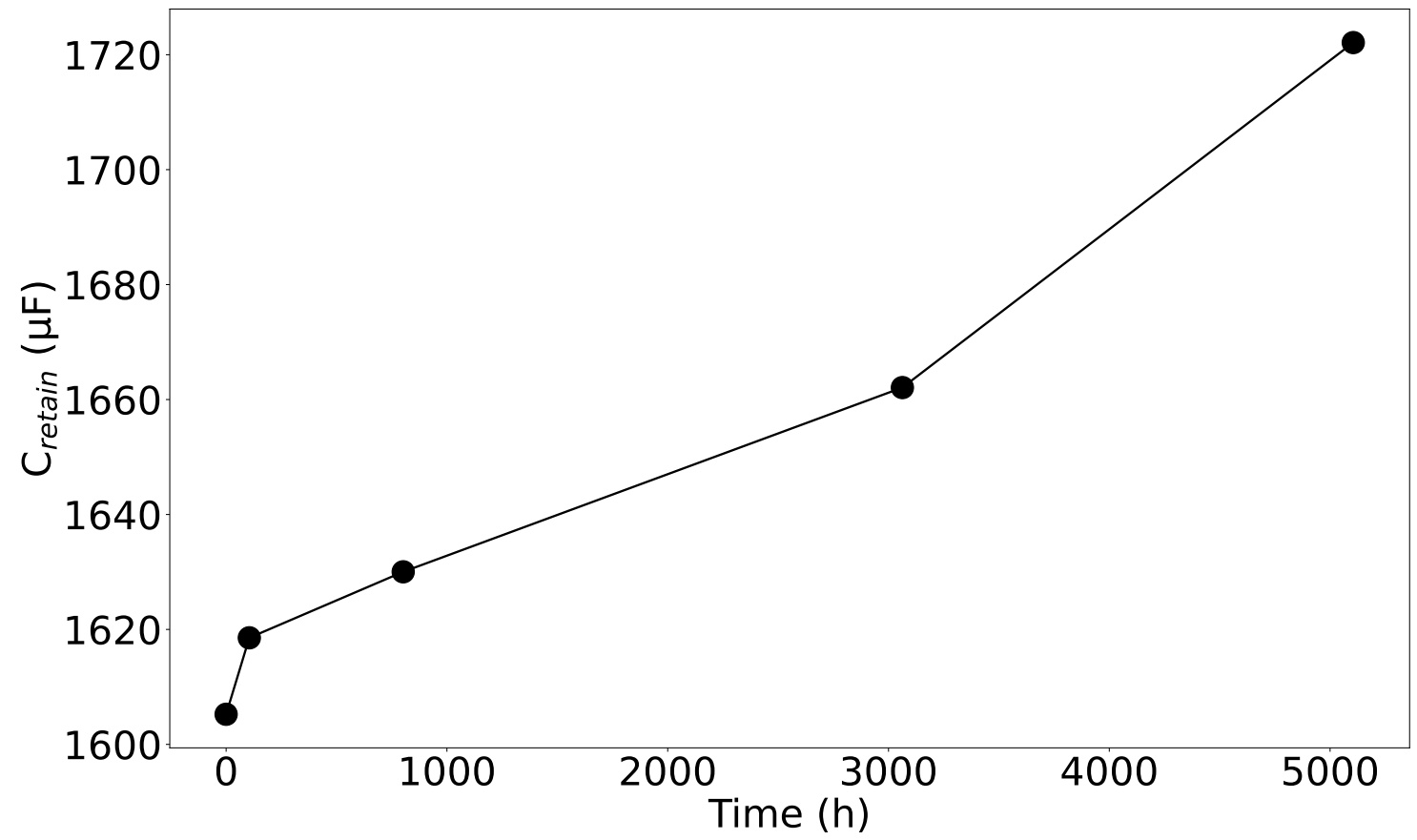

(a) $C_{rem}$

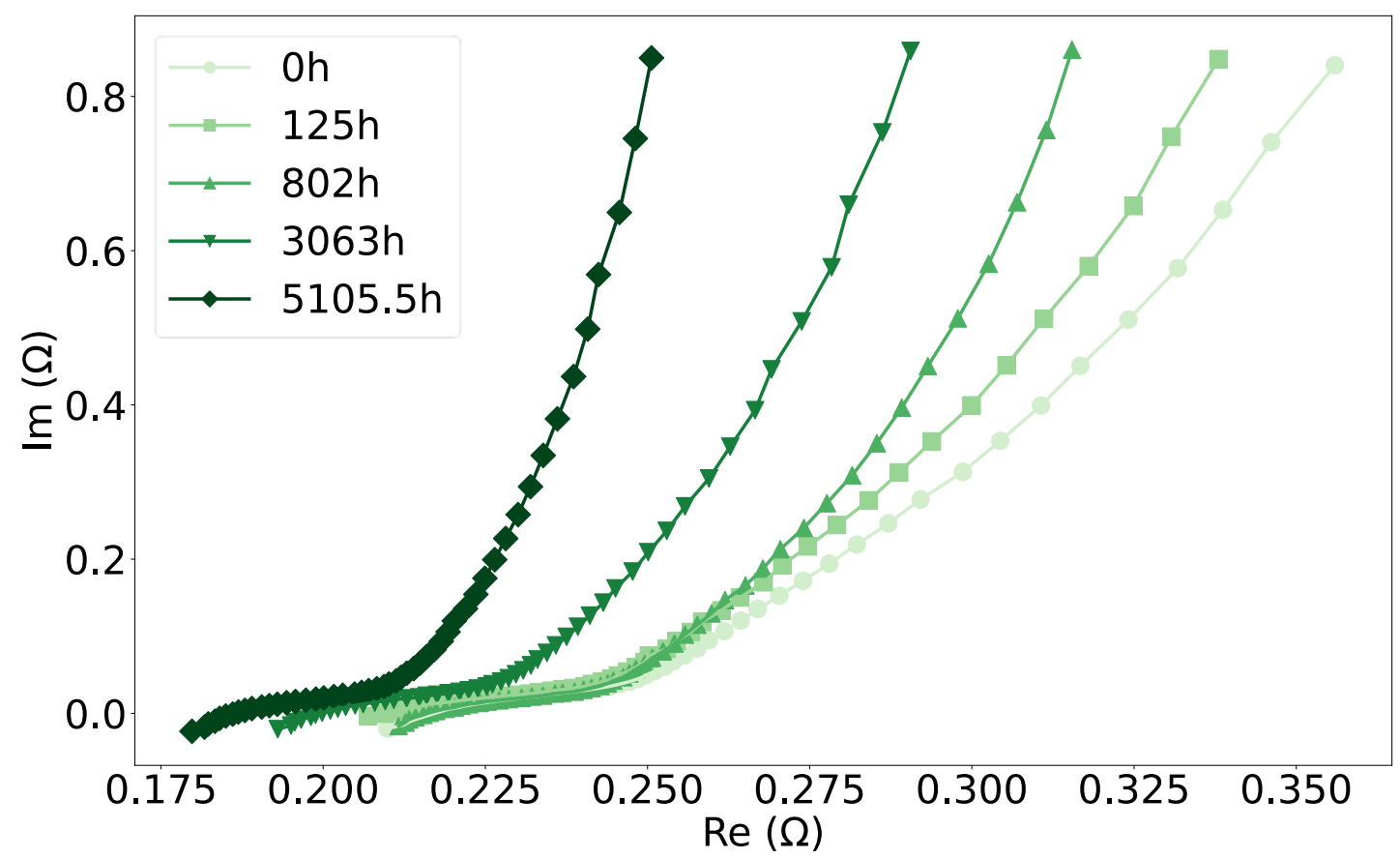

(b) EIS

**Supplementary Figure 3** Remaining capacitances and EIS of the representative capacitor (numbered as ES10C4 in the test set of Dataset 3) measured at different test stages.

## Supplementary Note 1

The preset test conditions corresponding to the EIS measurement include the PEMFC coolant outlet temperature ($T_{out}$), cathode inlet air humidity ($H_{ca,air}$), anode inlet hydrogen humidity ($H_{an,H2}$), cathode inlet air pressure ($P_{ca,air}$), anode inlet hydrogen pressure ($P_{an,H2}$), cathode inlet air flow ($F_{ca,air}$), anode inlet hydrogen flow ($F_{an,H2}$), load current ($I_{load}$), and injection current amplitude ($I_{amp}$), etc. In Dataset 1, $T_{out}$ is 80 °C, $H_{ca,air}$ is 100%, $H_{an,H2}$ is 100%, $P_{ca,air}$ is 2 bara, $P_{an,H2}$ is 2 bara, $F_{ca,air}$ is 5 NLPM, $F_{an,H2}$ is 2 NLPM, $I_{load}$ is 1.0 A/$cm^2$, and $I_{amp}$ is 0.1 A/$cm^2$.

As shown in **Supplementary Figure 4**, based on the Re/f curves in the training set of Dataset 1, the k-means clustering algorithm is utilized to extract two preset frequencies from the Re/f curve containing 41 frequencies. First, the medium frequency interval (1 Hz~100 Hz) and high frequency interval (100 Hz~10K Hz) in the Re/f curve are extracted. Each Re/f data point contains a frequency value and a corresponding Re value. First, the number of clusters $n$ is set to 2, and the *k-means++* algorithm is utilized to randomly select two data points on the Re/f curve as the initial cluster centers. Then, the k-means clustering algorithm is utilized to output two clusters of medium frequency and high frequency. Next, the mean of the frequency values in the medium frequency cluster is calculated, and the actual frequency value closest to the frequency mean is selected as the central frequency of the intermediate frequency cluster. Similarly, the central frequency of the high frequency cluster can be extracted. These two central frequencies are used as a central frequency combination. Finally, the number of occurrences of different central frequency combinations in all Re/f curves is counted, and the central frequency combination with the largest number of occurrences is selected as the two preset frequencies.

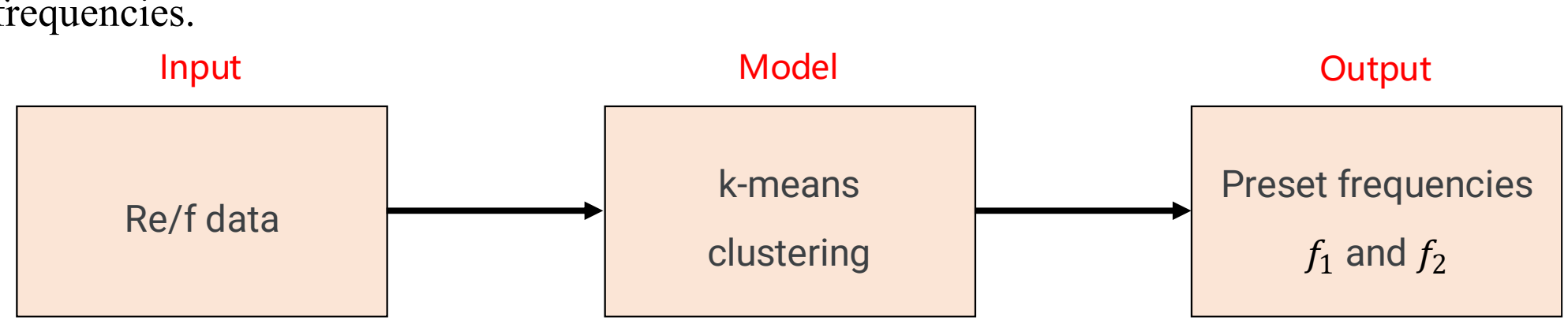

**Supplementary Figure 4** Selection method of two preset frequencies $f_1$ and $f_2$.

## Supplementary Note 2

In Dataset 1, the EIS prediction model uses the Random Forest (RF) model and is trained using the data in the training set. The input of the EIS prediction model is $Re_1$, $Re_2$, $Im_1$, and $Im_2$, and the output is 82 impedances (real and imaginary) corresponding to 41 frequencies in the mid-high frequency range (1 Hz~10k Hz), as shown in **Supplementary Figure 5**.

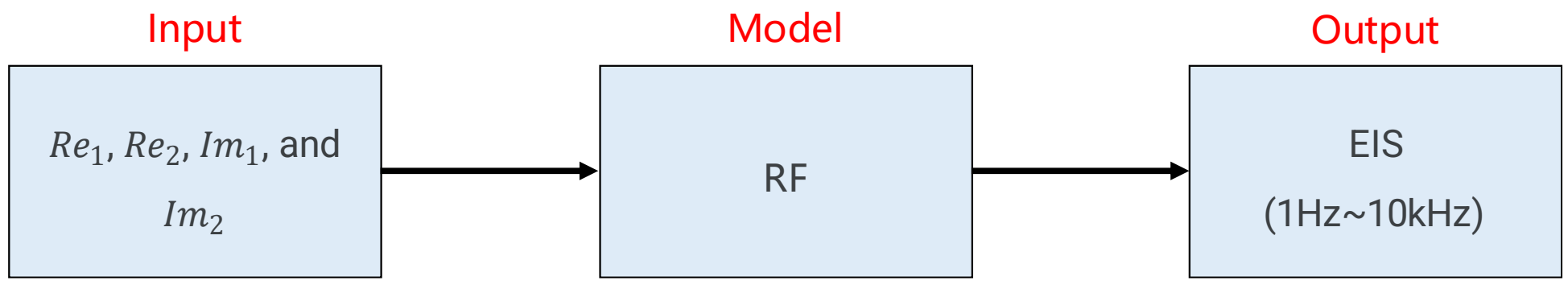

**Supplementary Figure 5** Prediction method for EIS (1 Hz~10k Hz).

For Dataset 1, the 41 frequencies in the mid-high frequency range (1 Hz~10k Hz) are shown in **Supplementary Table 1**.

**Supplementary Table 1** 41 frequencies in the mid-high frequency range (1 Hz~10k Hz).

| No. | Frequencies | No. | Frequencies | No. | Frequencies |
|---|---|---|---|---|---|
| 1 | 1.0000 | 16 | 31.6230 | 31 | 1000.0000 |
| 2 | 1.2589 | 17 | 39.8110 | 32 | 1258.9000 |
| 3 | 1.5849 | 18 | 50.1190 | 33 | 1584.9000 |
| 4 | 1.9953 | 19 | 63.0960 | 34 | 1995.3000 |
| 5 | 2.5119 | 20 | 79.4330 | 35 | 2511.9000 |
| 6 | 3.1623 | 21 | 100.0000 | 36 | 3162.3000 |
| 7 | 3.9811 | 22 | 125.8900 | 37 | 3981.1000 |
| 8 | 5.0119 | 23 | 158.4900 | 38 | 5011.9000 |
| 9 | 6.3096 | 24 | 199.5300 | 39 | 6309.6000 |
| 10 | 7.9433 | 25 | 251.1900 | 40 | 7943.3000 |
| 11 | 10.0000 | 26 | 316.2300 | 41 | 10000.0000 |
| 12 | 12.5890 | 27 | 398.1100 | | |
| 13 | 15.8490 | 28 | 501.1900 | | |
| 14 | 19.9530 | 29 | 630.9600 | | |
| 15 | 25.1190 | 30 | 794.3300 | | |

## Supplementary Note 3

In Dataset 1, the preset test conditions corresponding to the I-V curve measurement include $T_{out}$, $H_{ca,air}$, $H_{an,H2}$, $P_{ca,air}$, $P_{an,H2}$, $F_{ca,air}$, $F_{an,H2}$, voltage measurement range ($[V_1, V_2]$), voltage measurement interval ($V_r$), single voltage measurement holding time ($t_V$), etc. In Dataset 1, $T_{out}$ is 80 °C, $H_{ca,air}$ is 100%, $H_{an,H2}$ is 100%, $P_{ca,air}$ is 2 bara, $P_{an,H2}$ is 2 bara, $F_{ca,air}$ is 5 NLPM, $F_{an,H2}$ is 2 NLPM, $[V_1, V_2]$ is [0.2 V, 0.95 V], $V_r$ is 0.04 V, and $t_V$ is 5 min.

The preset test conditions corresponding to the CV curve measurement include $T_{out}$, cathode inlet gas humidity ($H_{ca,gas}$), $H_{an,H2}$, cathode inlet gas pressure ($P_{ca,gas}$), $P_{an,H2}$, cathode inlet gas flow ($F_{ca,gas}$), $F_{an,H2}$, CV cycle number ($n_{CV}$), voltage scan rate ($S_V$), voltage scan range ($[V_{s1}, V_{s2}]$), etc. In Dataset 1, $T_{out}$ is 80 °C, $H_{ca,gas}$ is 100%, $H_{an,H2}$ is 100%, $P_{ca,gas}$ is 2 bara, $P_{an,H2}$ is 2 bara, $F_{ca,gas}$ is 0 NLPM, $F_{an,H2}$ is 1 NLPM, $n_{CV}$ is 5, $S_V$ is 100 m$Vs^{-1}$, and $[V_{s1}, V_{s2}]$ is [0.05 V, 0.95 V].

The preset test conditions corresponding to the LSV curve include $T_{out}$, $H_{ca,air}$, $H_{an,H2}$, $P_{ca,air}$, $P_{an,H2}$, $F_{ca,air}$, $F_{an,H2}$, $S_V$, $[V_{s1}, V_{s2}]$, etc. In Dataset 1, $T_{out}$ is 80 °C, $H_{ca,air}$ is 100%, $H_{an,H2}$ is 100%, $P_{ca,air}$ is 2 bara, $P_{an,H2}$ is 2 bara, $F_{ca,air}$ is 1 NLPM, $F_{an,H2}$ is 1 NLPM, $S_V$ is 1 m$Vs^{-1}$,, and $[V_{s1}, V_{s2}]$ is [0.1 V, 0.5 V].

The I-V curve prediction model, CV curve prediction model, and LSV curve prediction model all use RF models. As shown in **Supplementary Figure 6**, in Dataset 1, the outputs of the I/V curve prediction model are 20 voltage values in the current density interval of [0 A$cm^{-2}$, 3.1 A$cm^{-2}$], and the interval is 0.16 A$cm^{-2}$. The outputs of the CV curve prediction model are 100 current density values in the voltage interval of [0.051 V, 0.40 V], and the interval is 0.004 V. The outputs of the LSV curve prediction model are 100 current density values in the voltage interval of [0.1 V, 0.5 V], and the interval is 0.004 V.

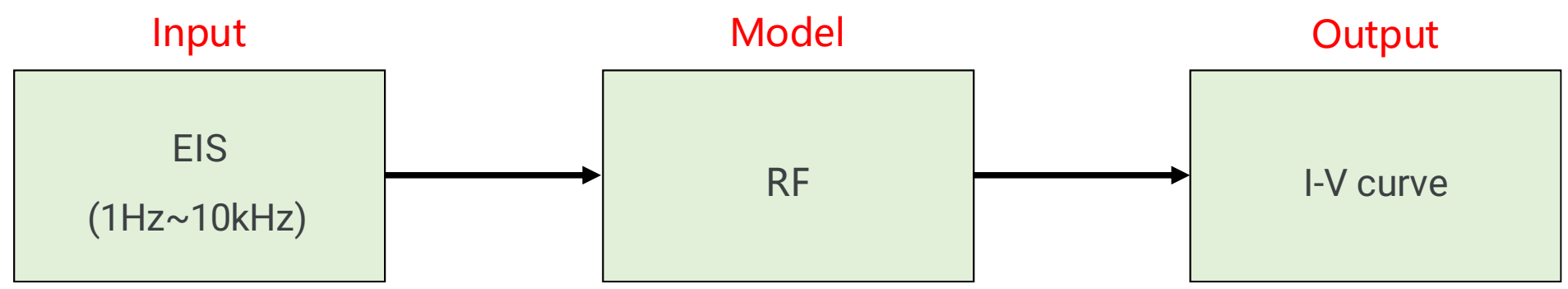

(a) Prediction method for I-V curve.

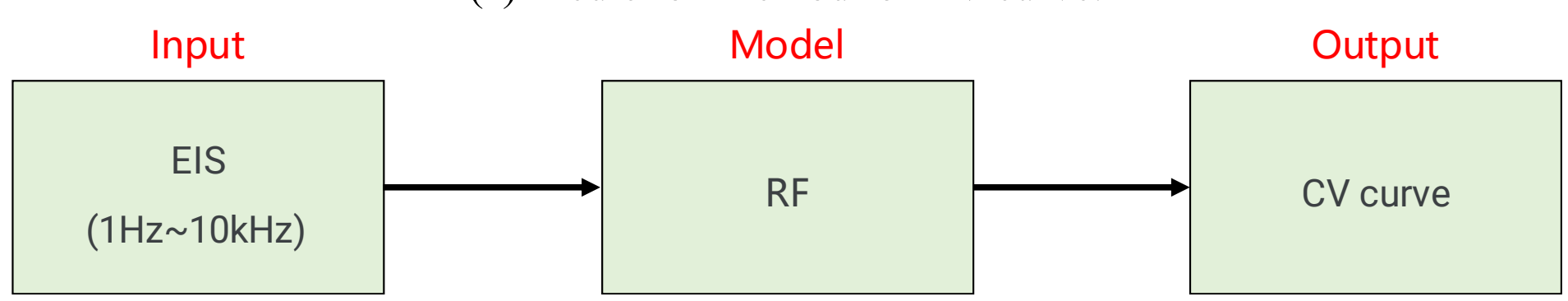

(b) Prediction method for CV curve.

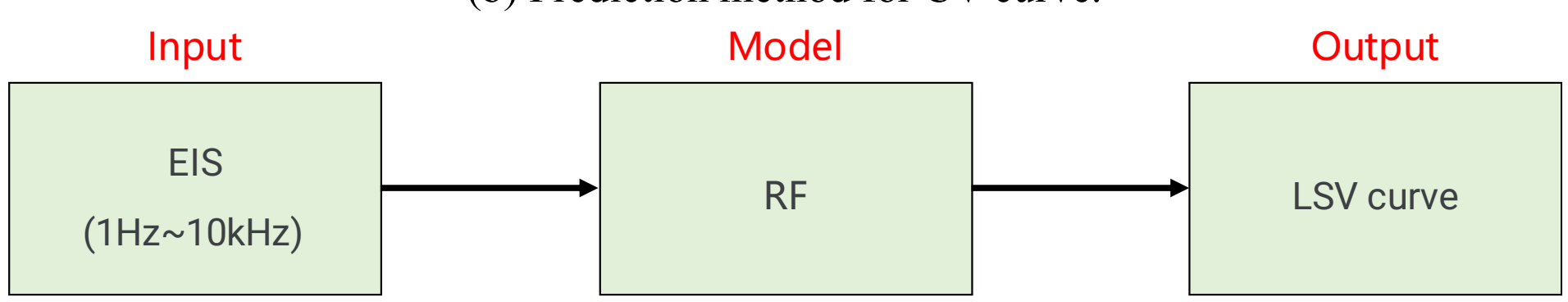

(c) Prediction method for LSV curve.

**Supplementary Figure 6** Prediction methods for different PCD.

## Supplementary Note 4

As shown in **Supplementary Figure 7**, in Dataset 1, the input of the $R_{O2,total}$ prediction model is the predicted/measured EIS (1 Hz~10k Hz), and the output is the predicted result of $R_{O2,total}$ . The input of the $I_{lim}$ prediction model is the predicted/measured I-V curve, and the output is the predicted $I_{lim}$. The input of the $ECSA$ prediction model is the predicted/measured CV curve, and the output is the predicted $ECSA$. The input of the $I_{cross}$ prediction model is the predicted/measured LSV curve, and the output is the predicted $I_{cross}$.

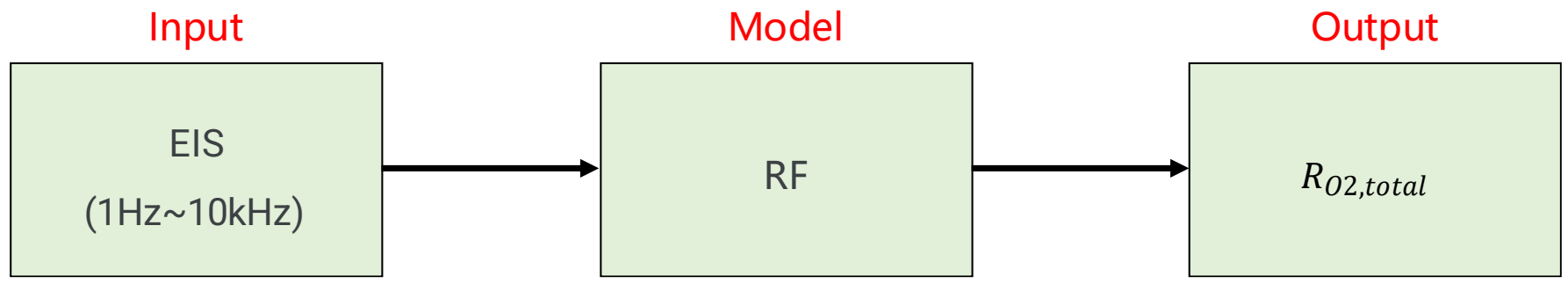

(a) Prediction method for $R_{O2,total}$.

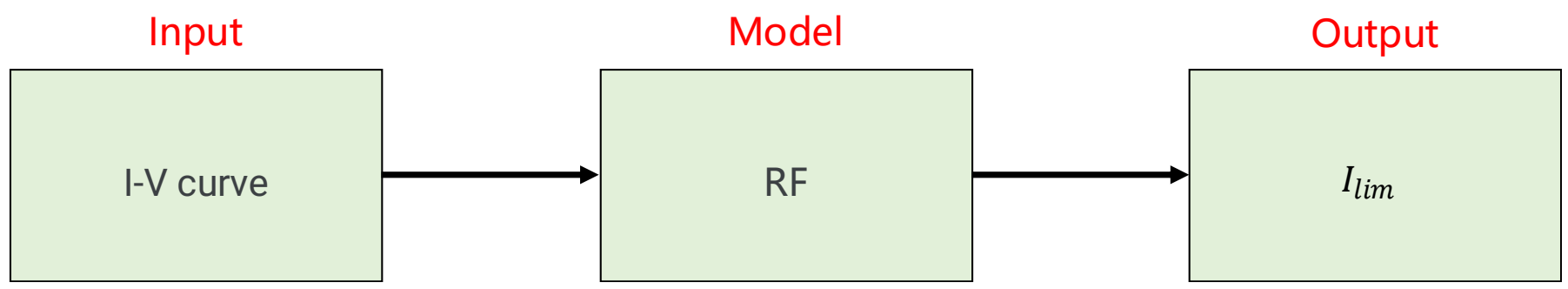

(b) Prediction method for $I_{lim}$.

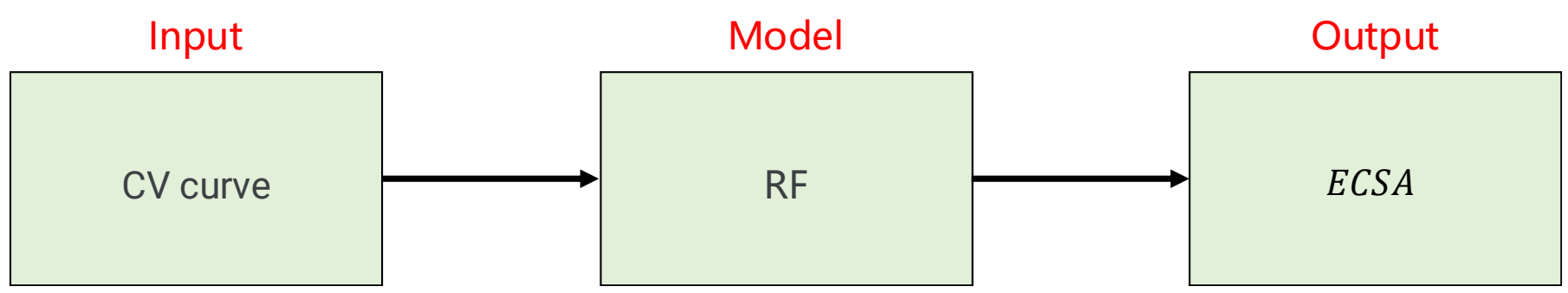

(c) Prediction method for $ECSA$.

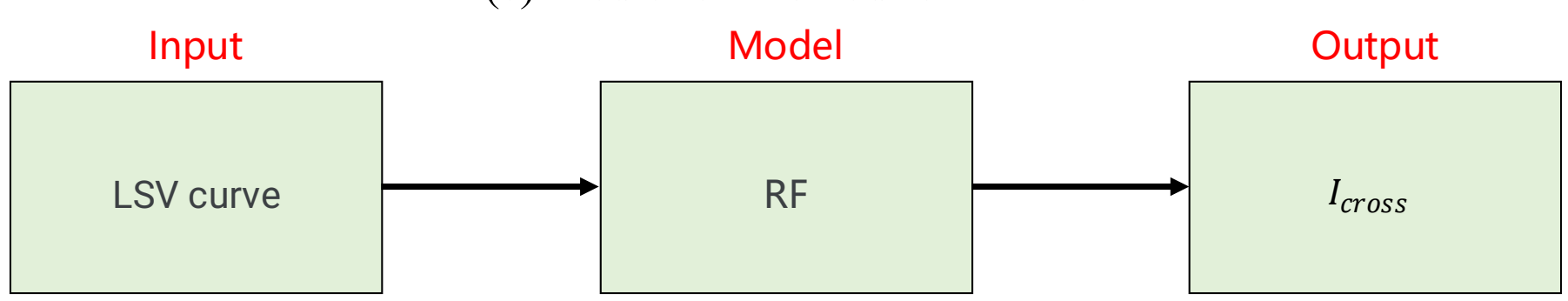

(d) Prediction method for $I_{cross}$.

**Supplementary Figure 7** Prediction methods for different aging indicators.

## Supplementary Note 5

The grid search method [1] is selected for hyperparameter optimization of the RF regression model [2]. The grid search method is a systematic hyperparameter optimization method that exhaustively searches all possible parameter combinations to find the best hyperparameters by defining a set of parameter grids. In this paper, the hyperparameter optimization of the RF regression model includes the number of trees (*n_estimators*), max depth (*max_depth*), minimum number of samples required to be at a leaf node (*min_samples_leaf*), the number of features to consider when looking for the best split (*max_features*), and the fraction of samples to be used for training each tree (*subsample*) [3]. Specifically, *n_estimators* has 6 values (50, 100, 150, 200, 250, and 300), *max_depth* has 6 values (5, 10, 15, 20, 25, and 30), *min_samples_leaf* has 5 values (1, 2, 3, 4, and 5), *max_features* has 6 values ('auto', 'sqrt', 0.33, 0.5, 0.75, and 1.0), and there are 1080 combinations of all hyperparameter values.

## Supplementary Note 6

Mean absolute error (MAE), mean absolute percentage error (MAPE), and root mean square error (RMSE) are chosen to evaluate prediction results of PCDP and LP-ALT methods.

MAE is defined as

$$\mathrm{MAE} = \frac{1}{n}\sum_{i=1}^{n} |y_i - \hat{y}_i| \tag{S1}$$

where $n$ is the number of samples, $y_i$ is the actual value of the $i$th sample, and $\hat{y}_i$ is the predicted value of the $i$th sample.

MAPE is defined as

$$\mathrm{MAPE} = \frac{100\%}{n}\sum_{i=1}^{n} \left|\frac{y_i - \hat{y}_i}{y_i}\right| \tag{S2}$$

RMSE is defined as

$$\mathrm{RMSE} = \sqrt{\frac{1}{n}\sum_{i=1}^{n} (y_i - \hat{y}_i)^2} \tag{S3}$$

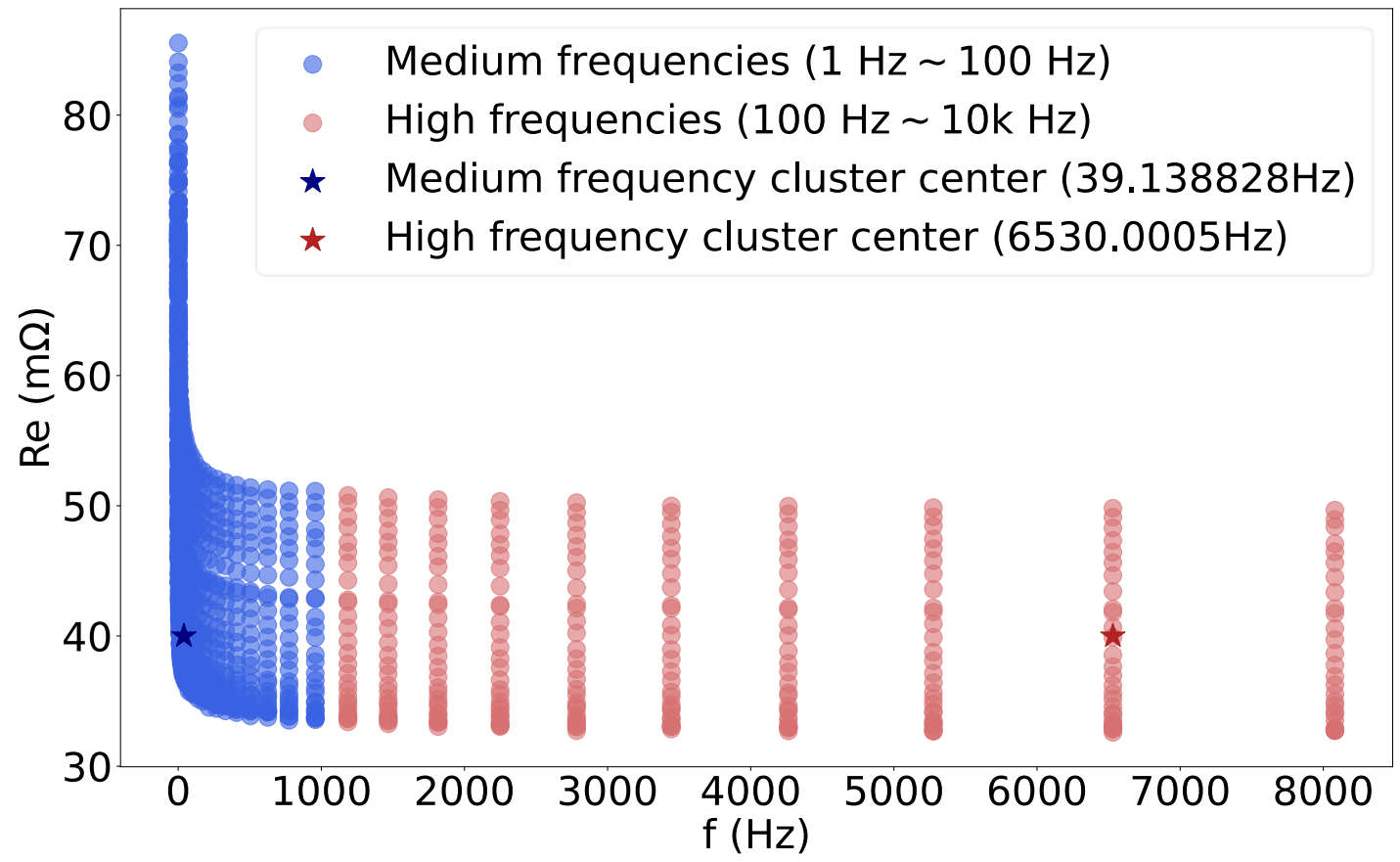

(a) The clustering results.

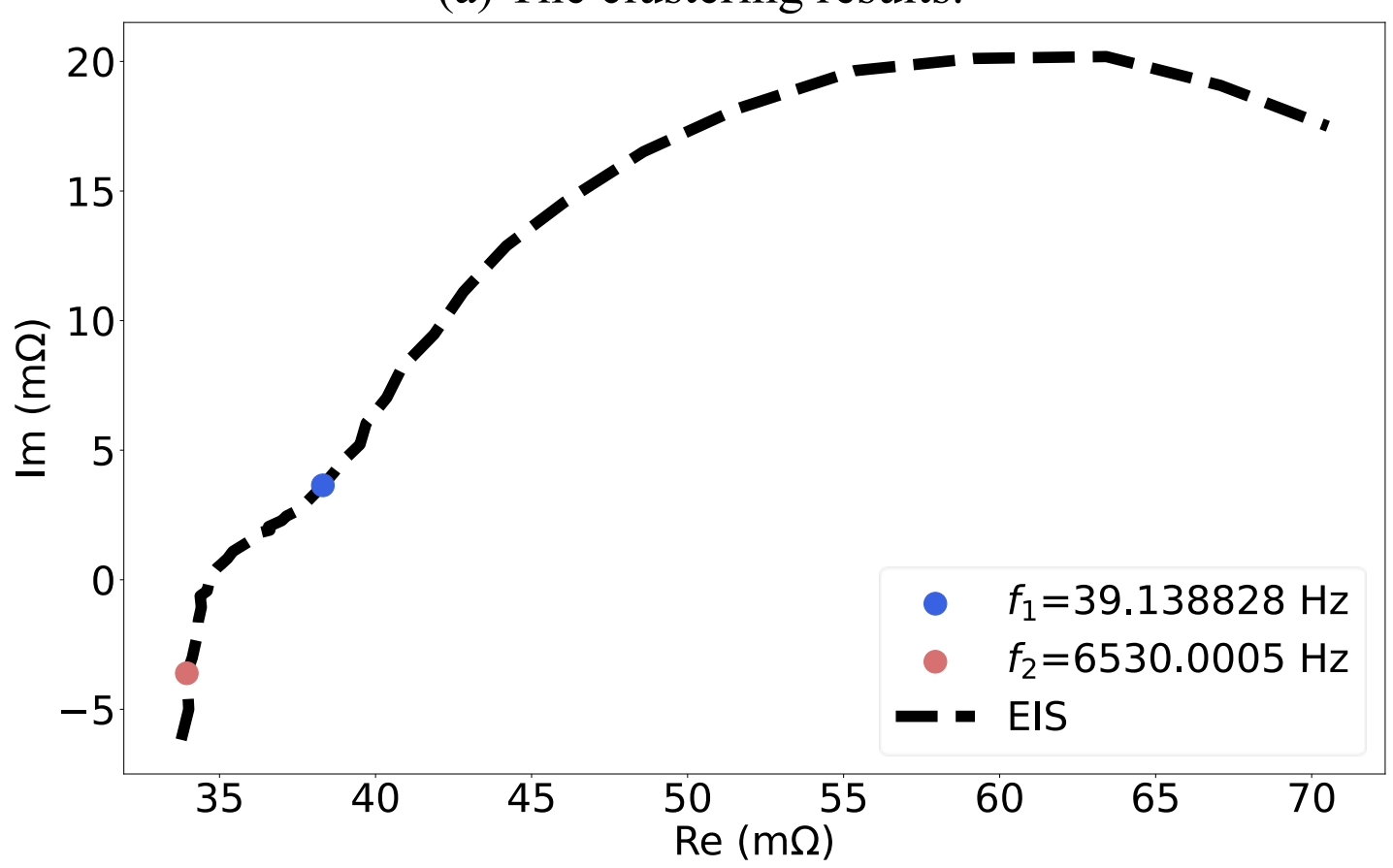

(b) Two preset frequencies $f_1$ and $f_2$ on the EIS.

**Supplementary Figure 5** Test results of Step 1 on the training set of Dataset 2. a. The clustering results in the medium frequency range (1 Hz~100 Hz) and the high frequency range (100 Hz~10k Hz) of Re/f curve. b. The positions of the two preset frequencies $f_1$ and $f_2$ on the EIS of the representative PCD measurement (numbered as 1th in the training set of Dataset 1).

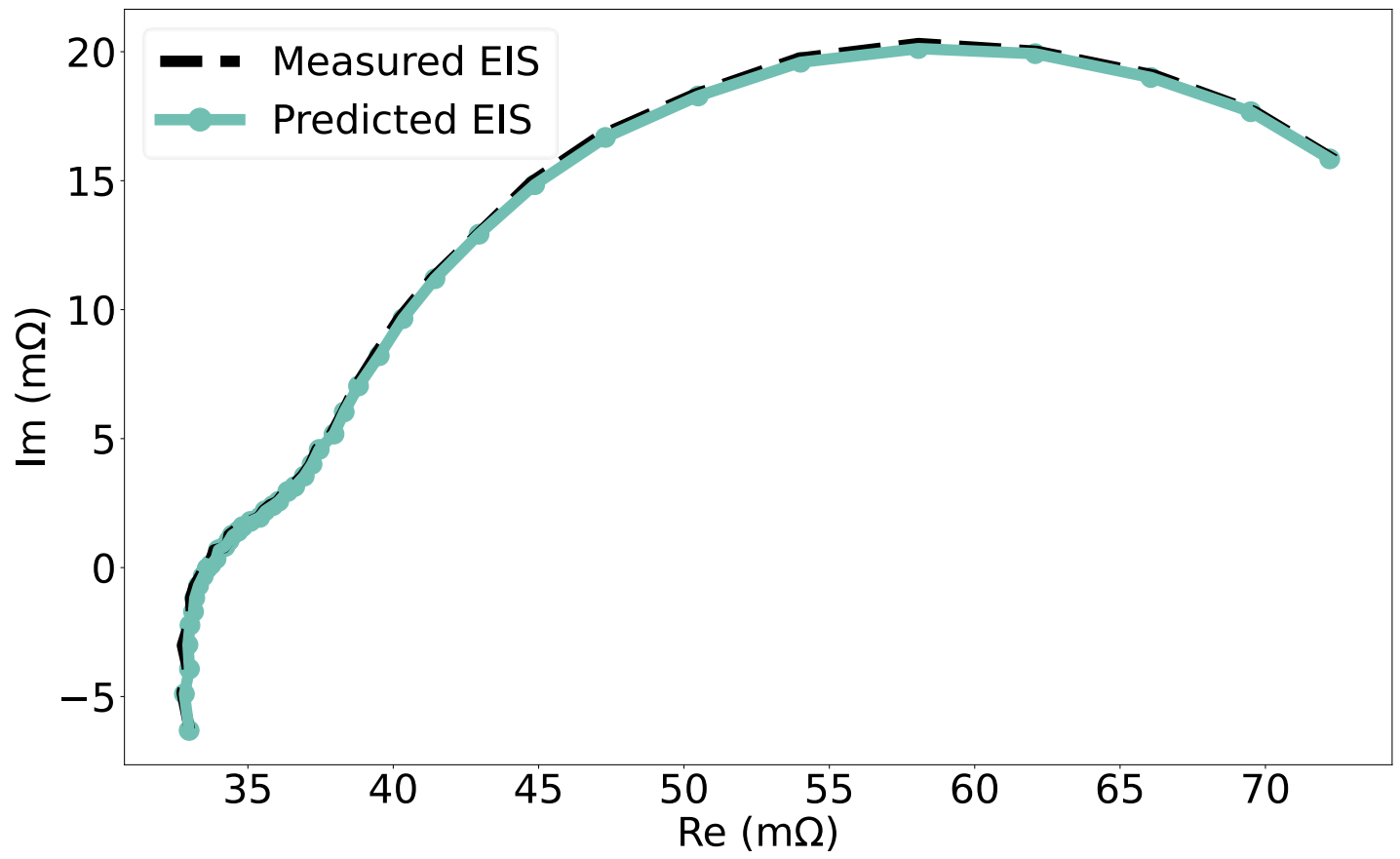

**Supplementary Figure 6** Test results of Step 2 on the representative PCD measurement (numbered as 4th in the test set of Dataset 2). The prediction results of EIS in the mid-high frequency range (1 Hz~10k Hz).

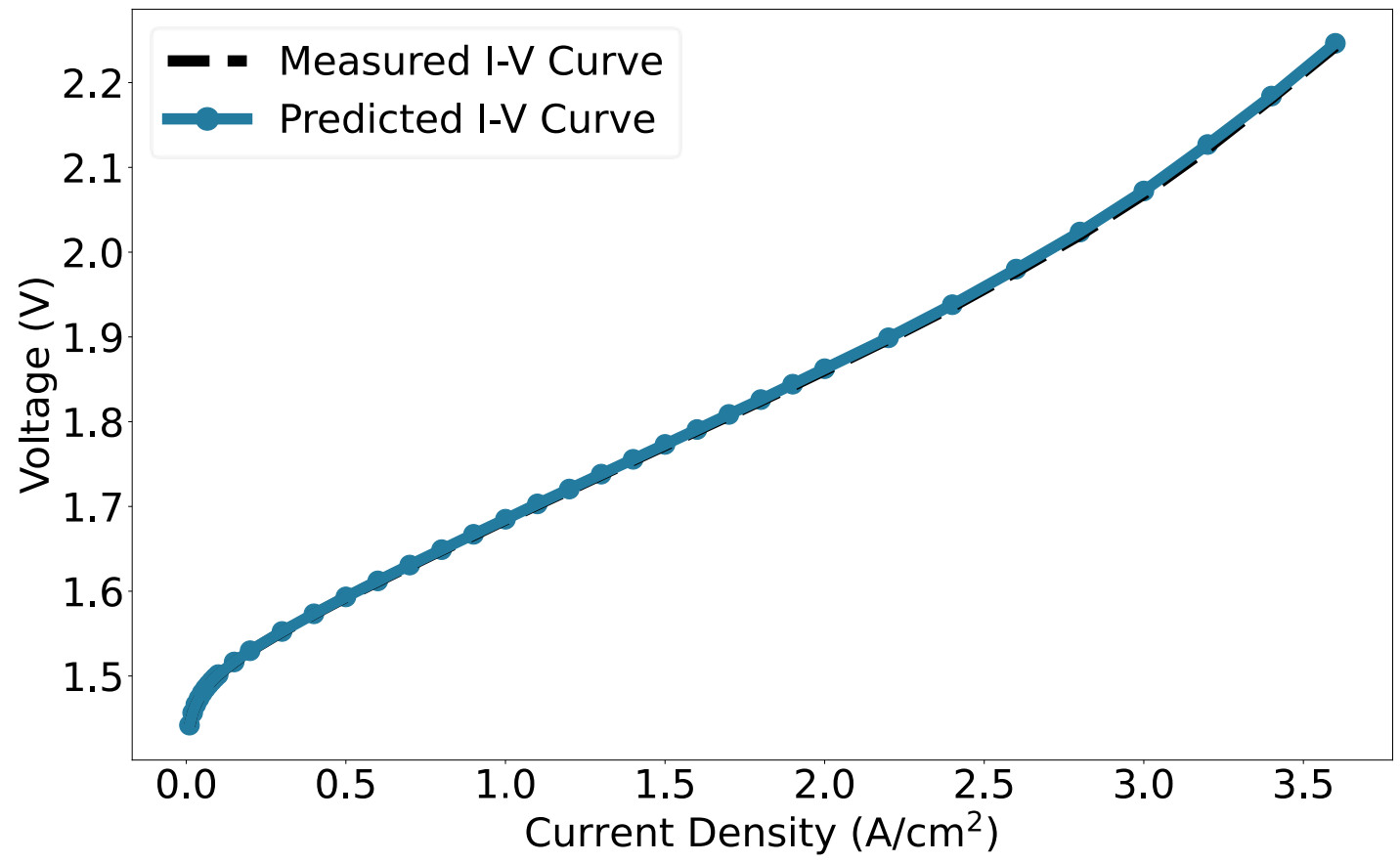

**Supplementary Figure 7** Test results of Step 3 on the representative PCD measurements (numbered as 4th in the test set of Dataset 2). The prediction results of I-V curve.

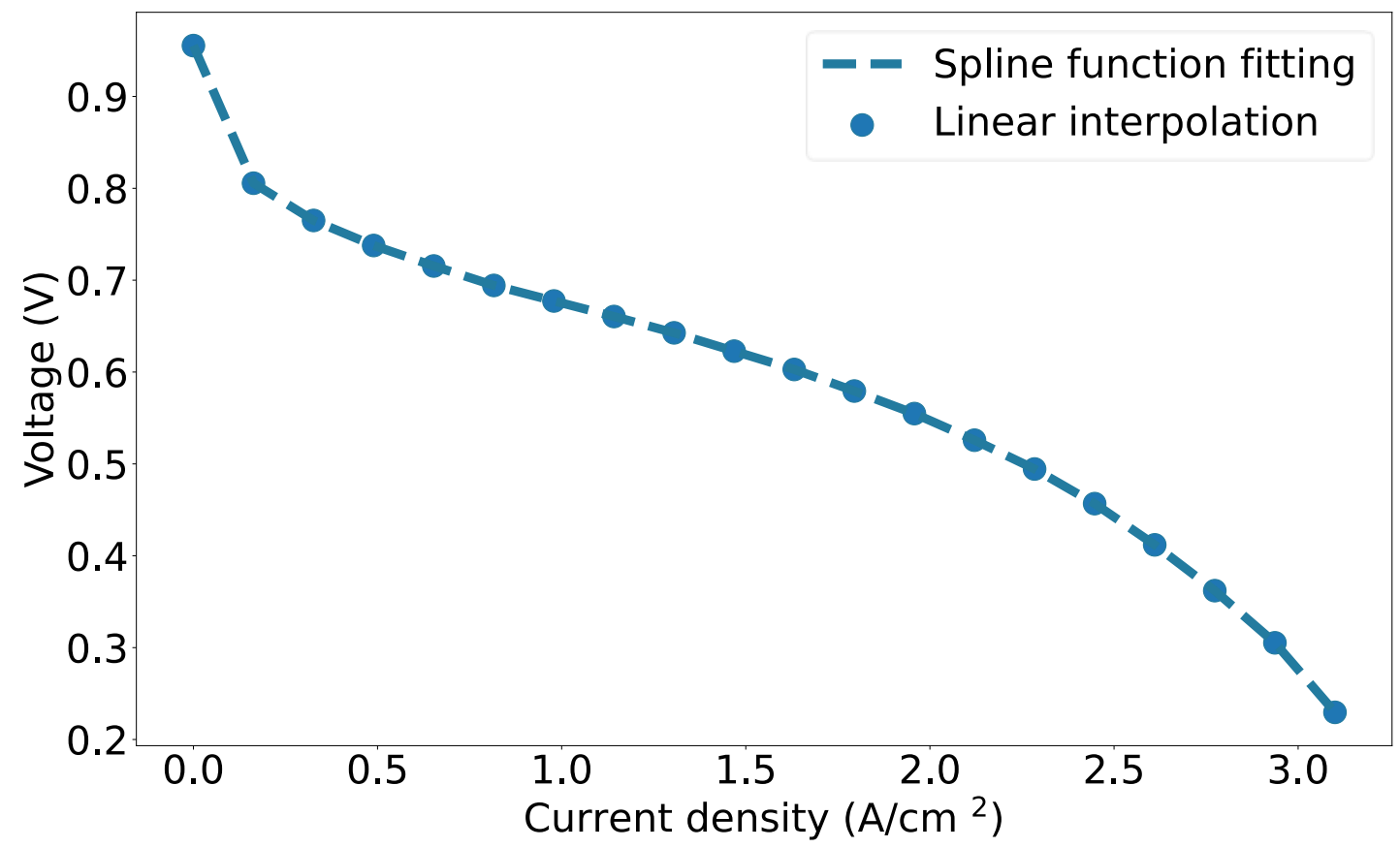

(a) Standardization of I-V curve from the representative PEMFC (numbered as 1 in Dataset 1).

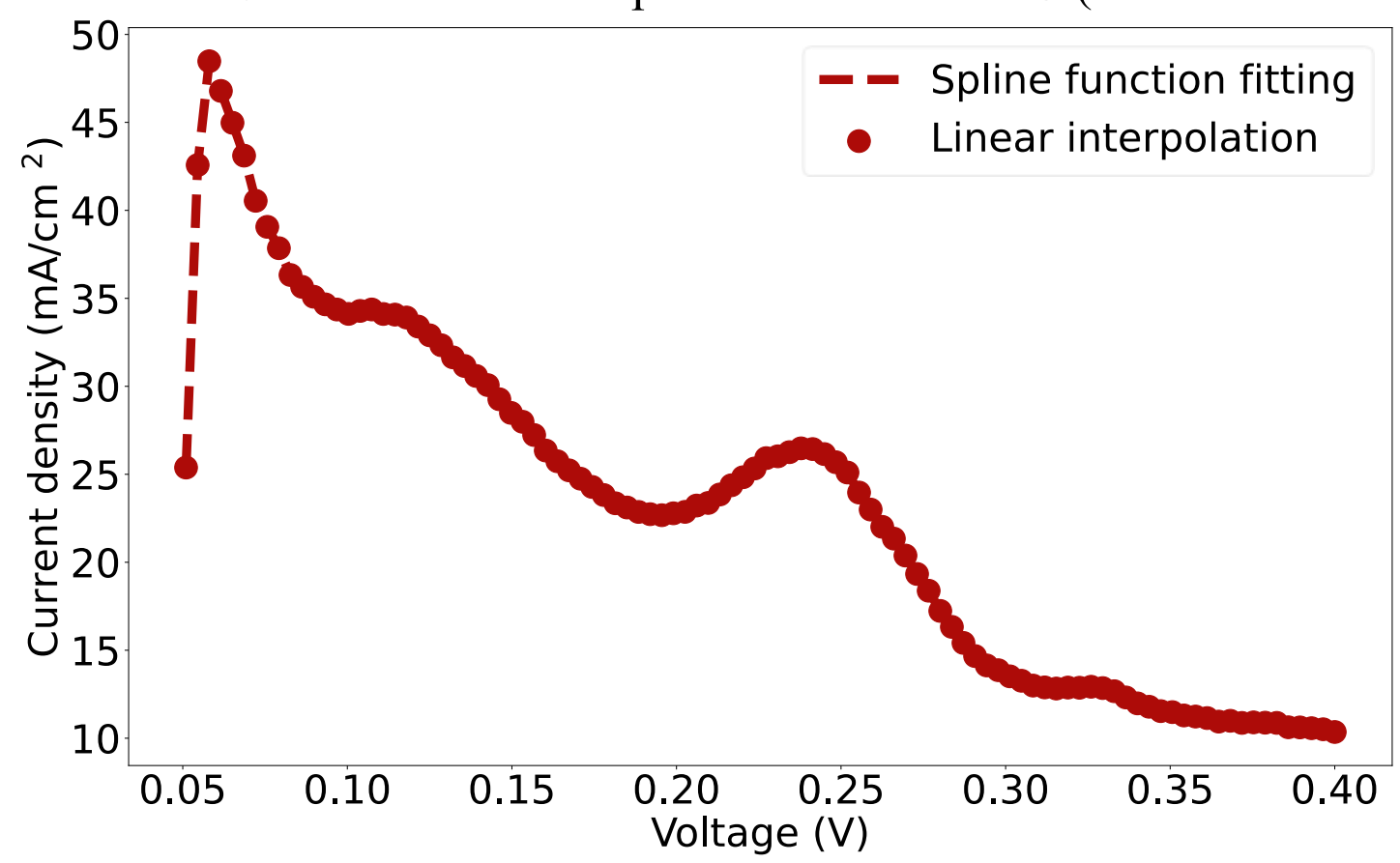

(b) Standardization of CV curve from the representative PEMFC (numbered as 1 in Dataset 1).

**Supplementary Figure 8** Standardization of different PCD.

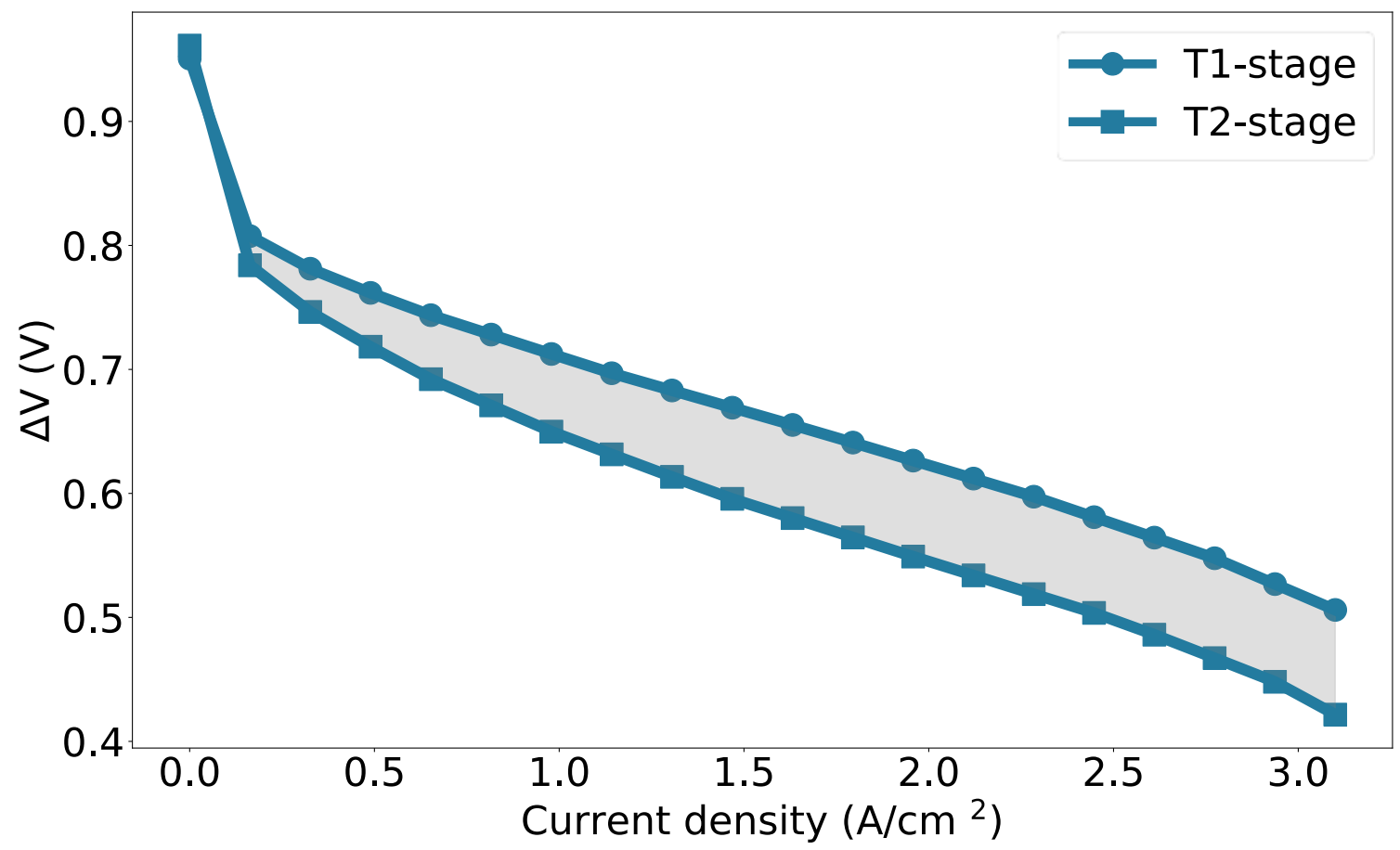

(a) Calculation method.

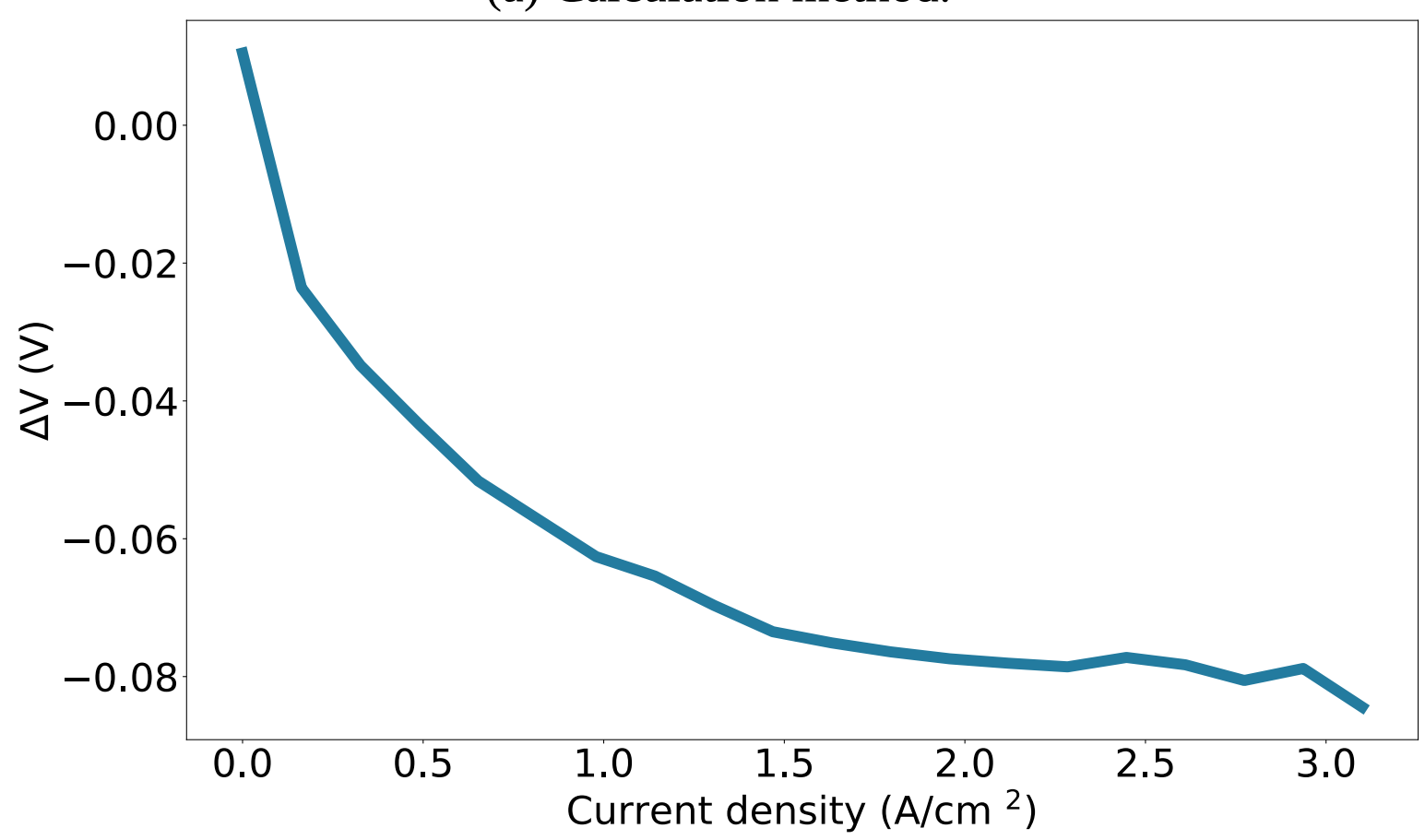

(b) Calculation results.

**Supplementary Figure 9** Calculation of the ΔV/I curve from the representative PEMFC (numbered as 1 in Dataset 1).

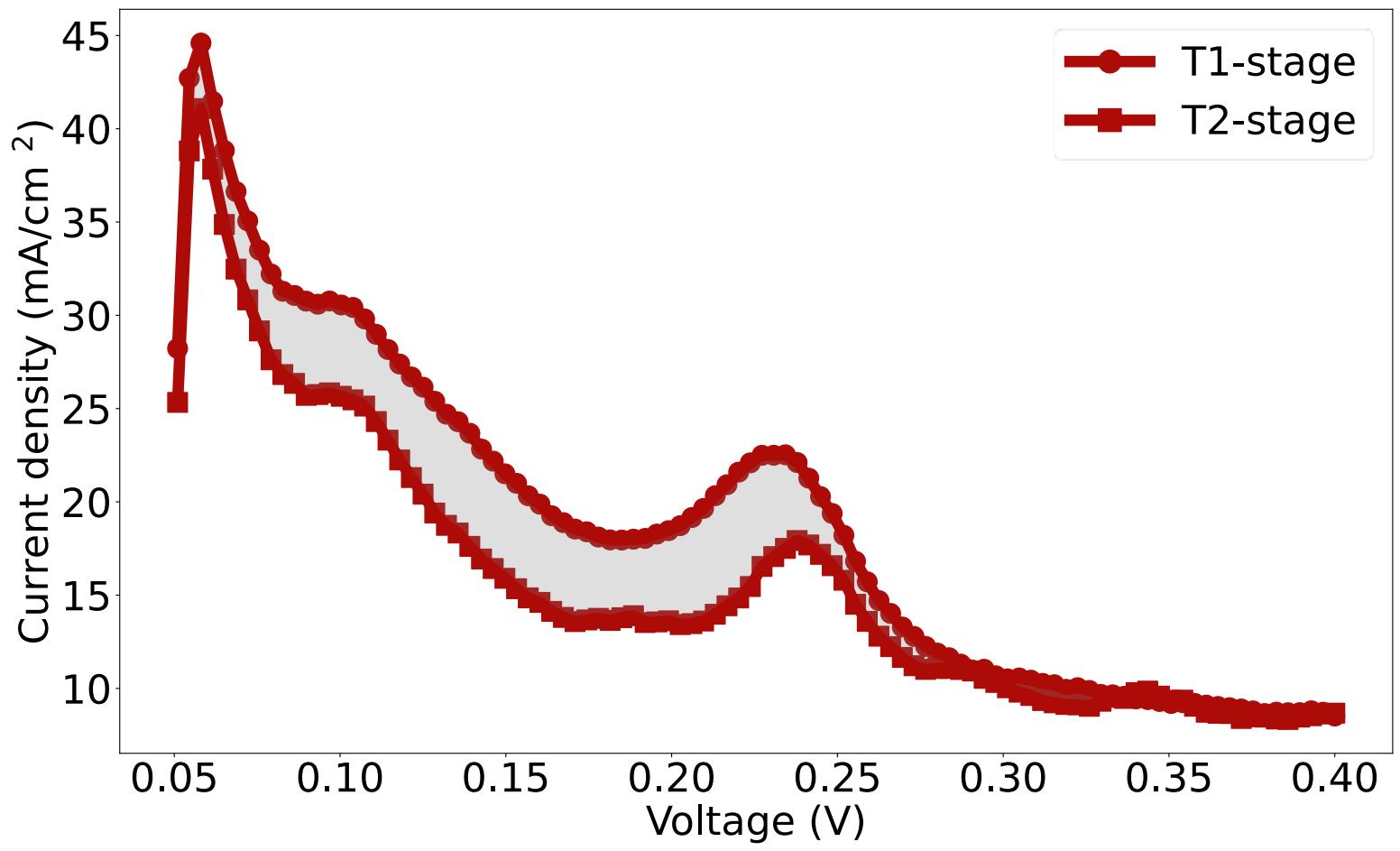

(a) Calculation method.

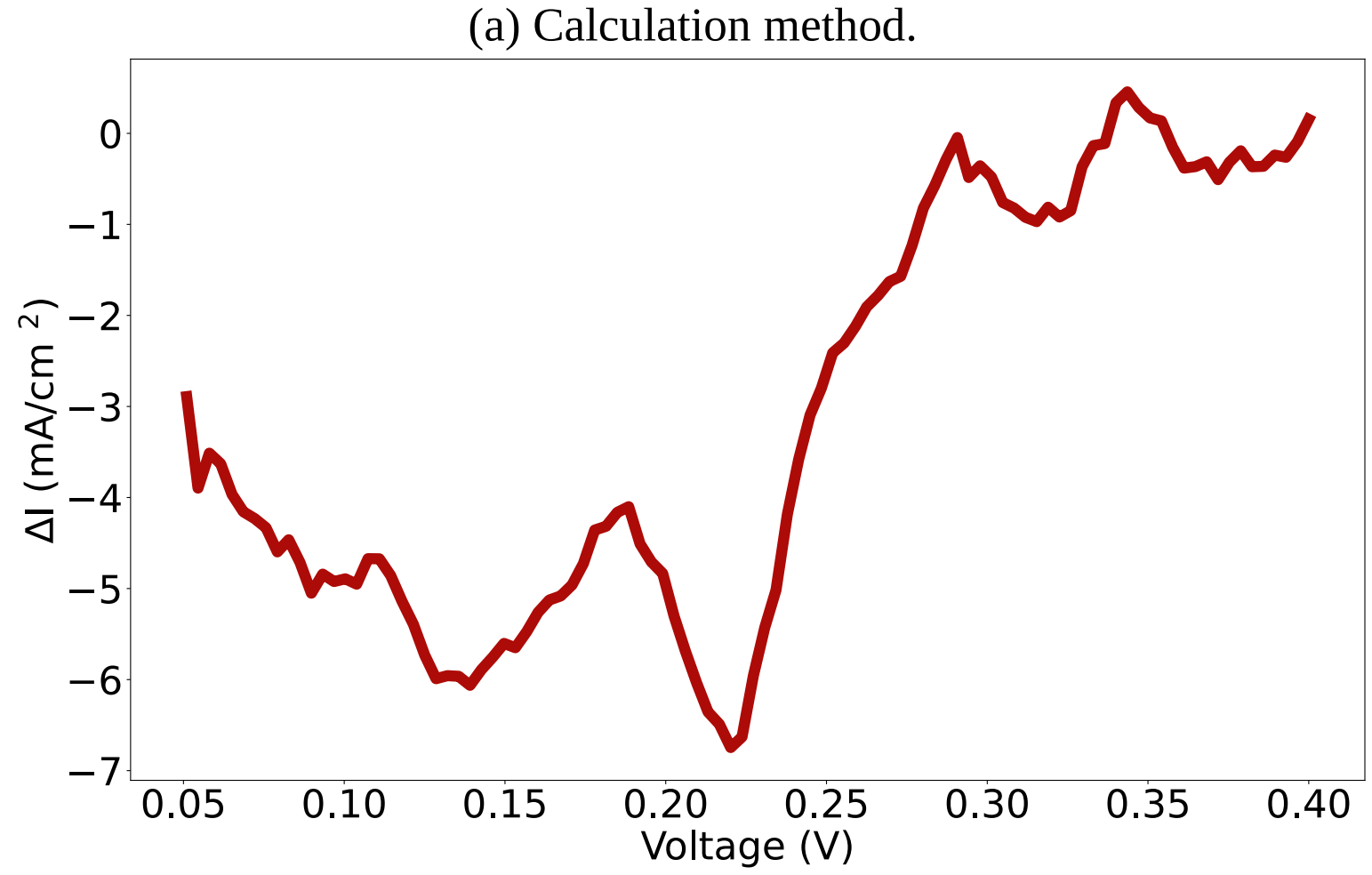

(b) Calculation results.

**Supplementary Figure 10** Calculation of the ΔI/V curve from the representative PEMFC (numbered as 1 in Dataset 1).

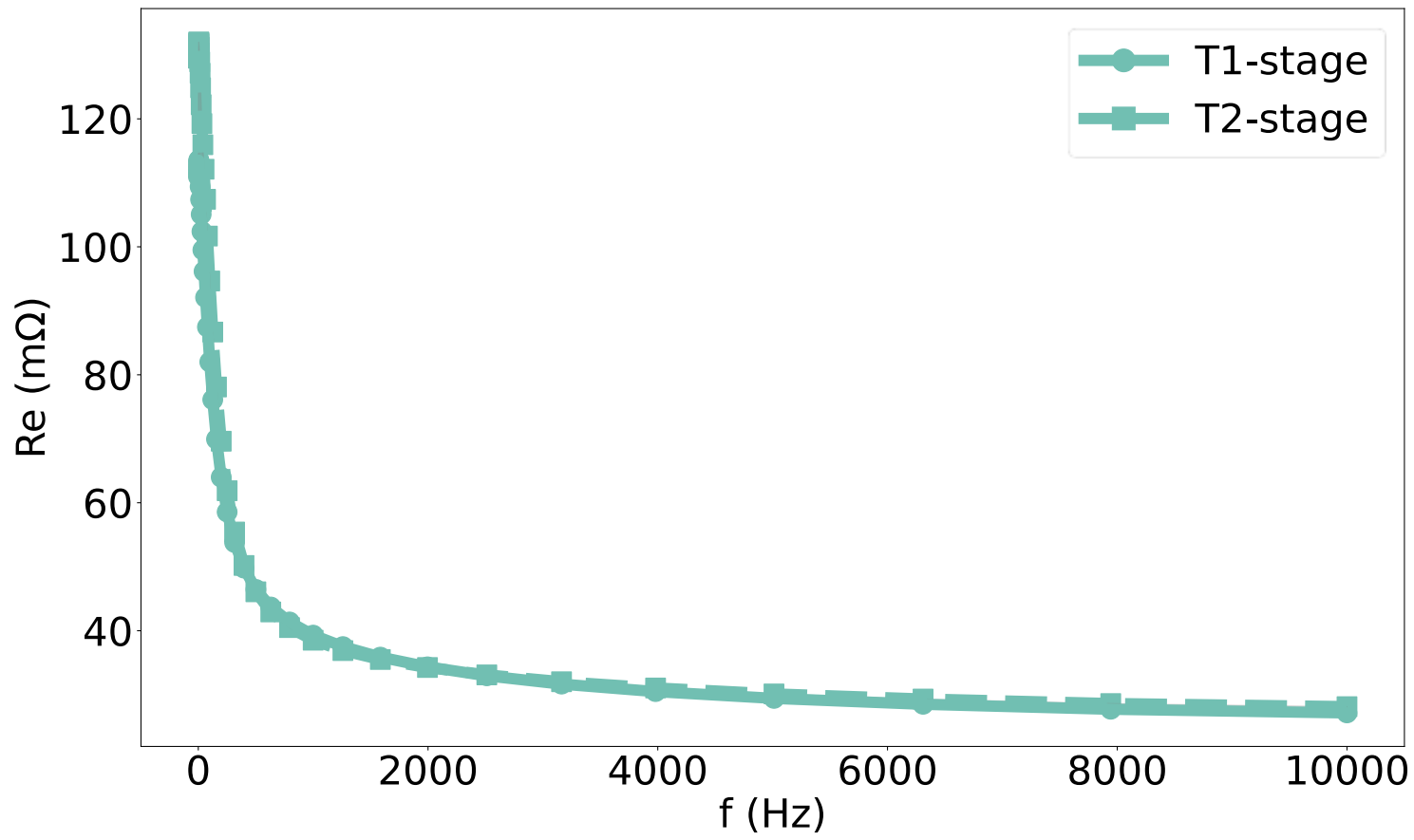

(a) Calculation method of ΔRe/f curve.

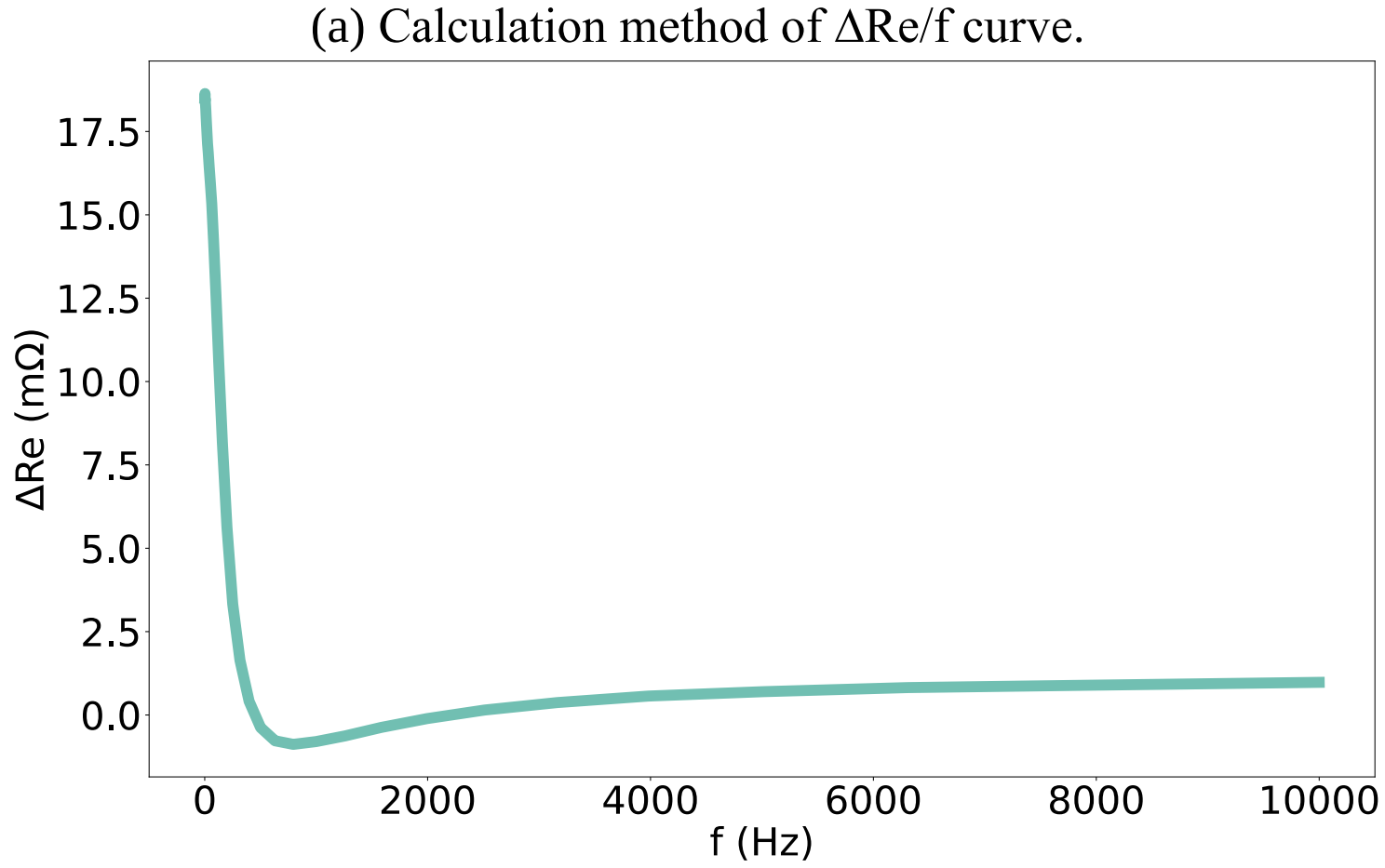

(b) Calculation results of ΔRe/f curve.

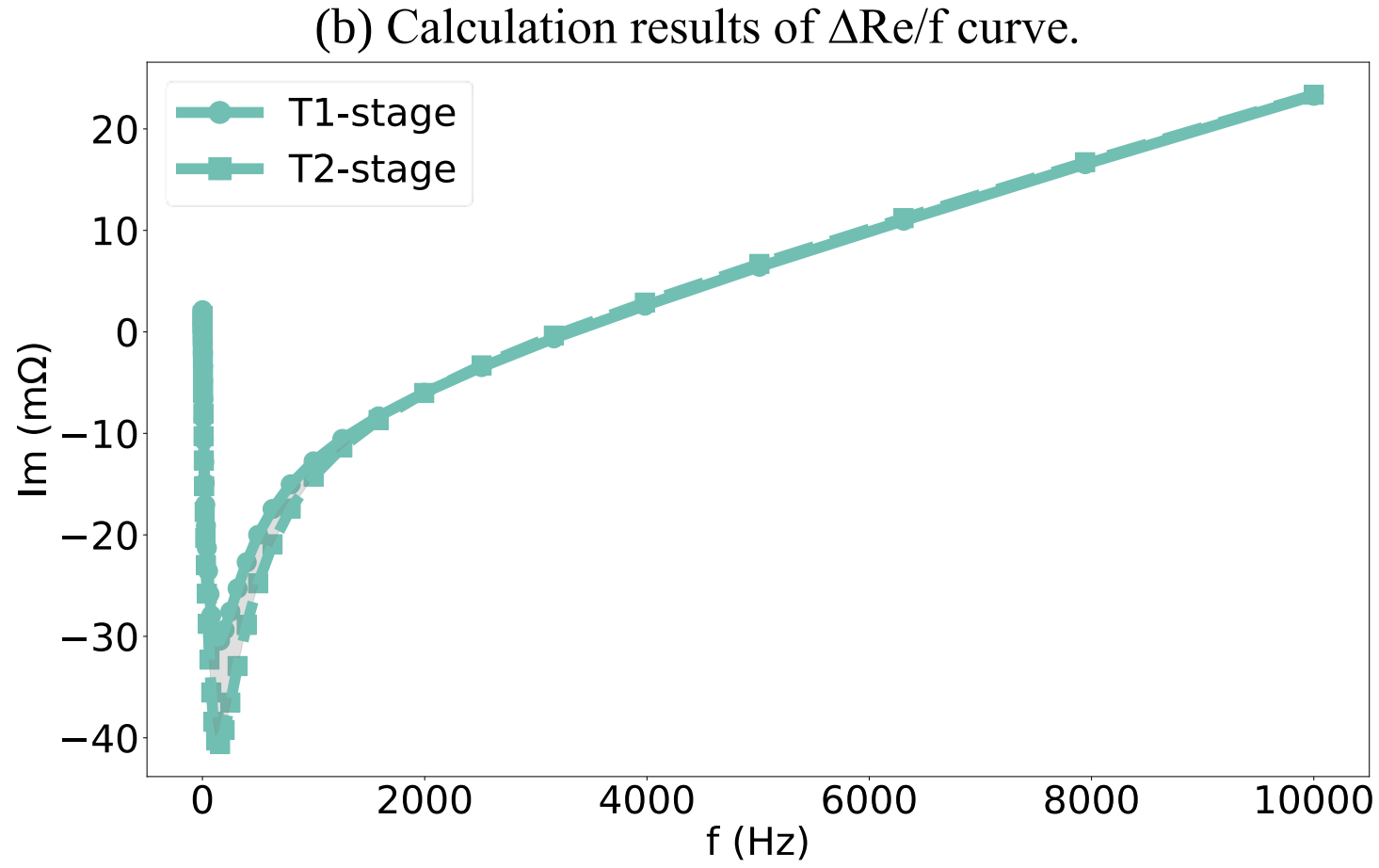

(c) Calculation method of ΔIm/f curve.

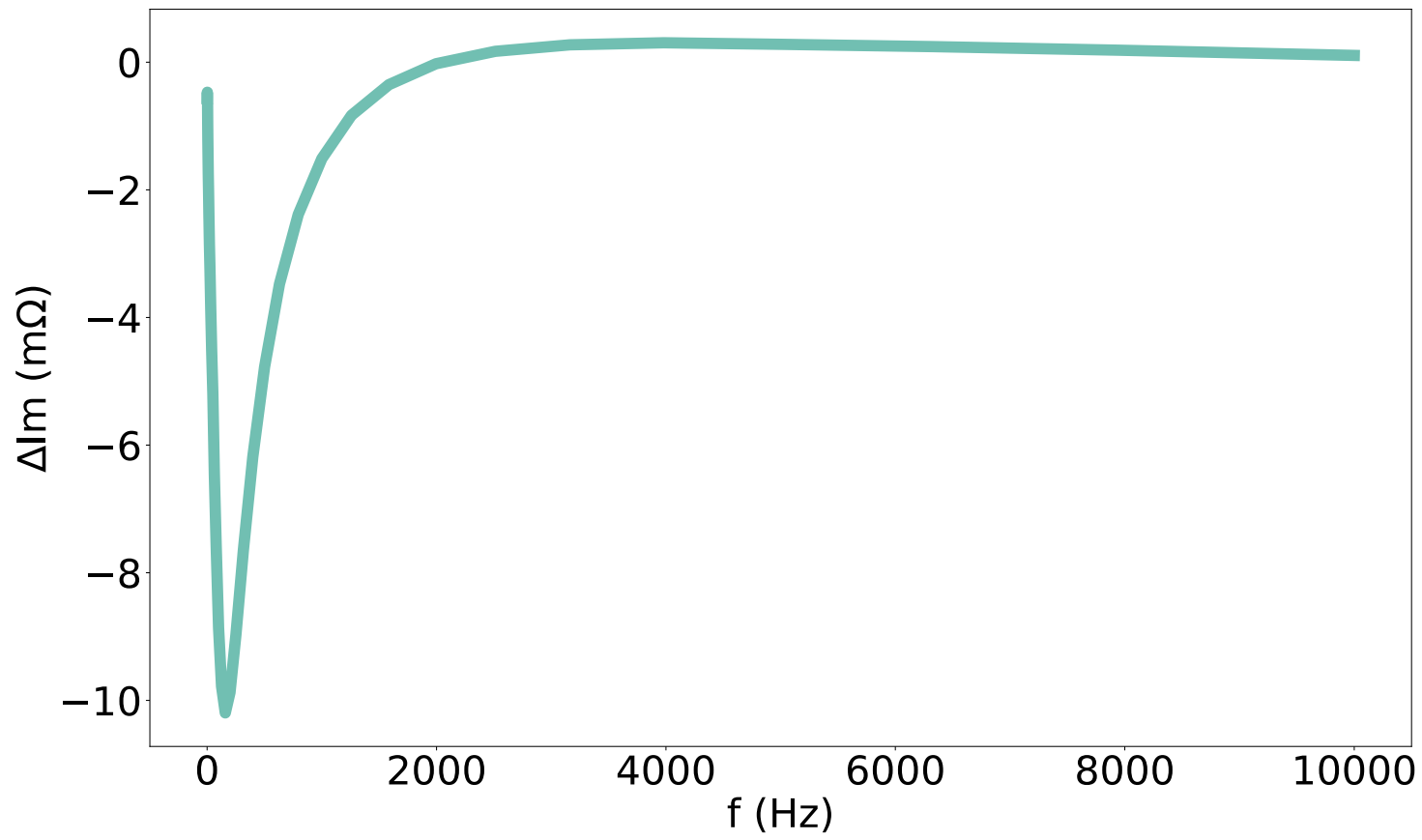

(d) Calculation results of ΔIm/f curve.

**Supplementary Figure 11** Calculation of the ΔRe/f curve and the ΔIm/f curve from the representative PEMFC (numbered as 1 in Dataset 1).

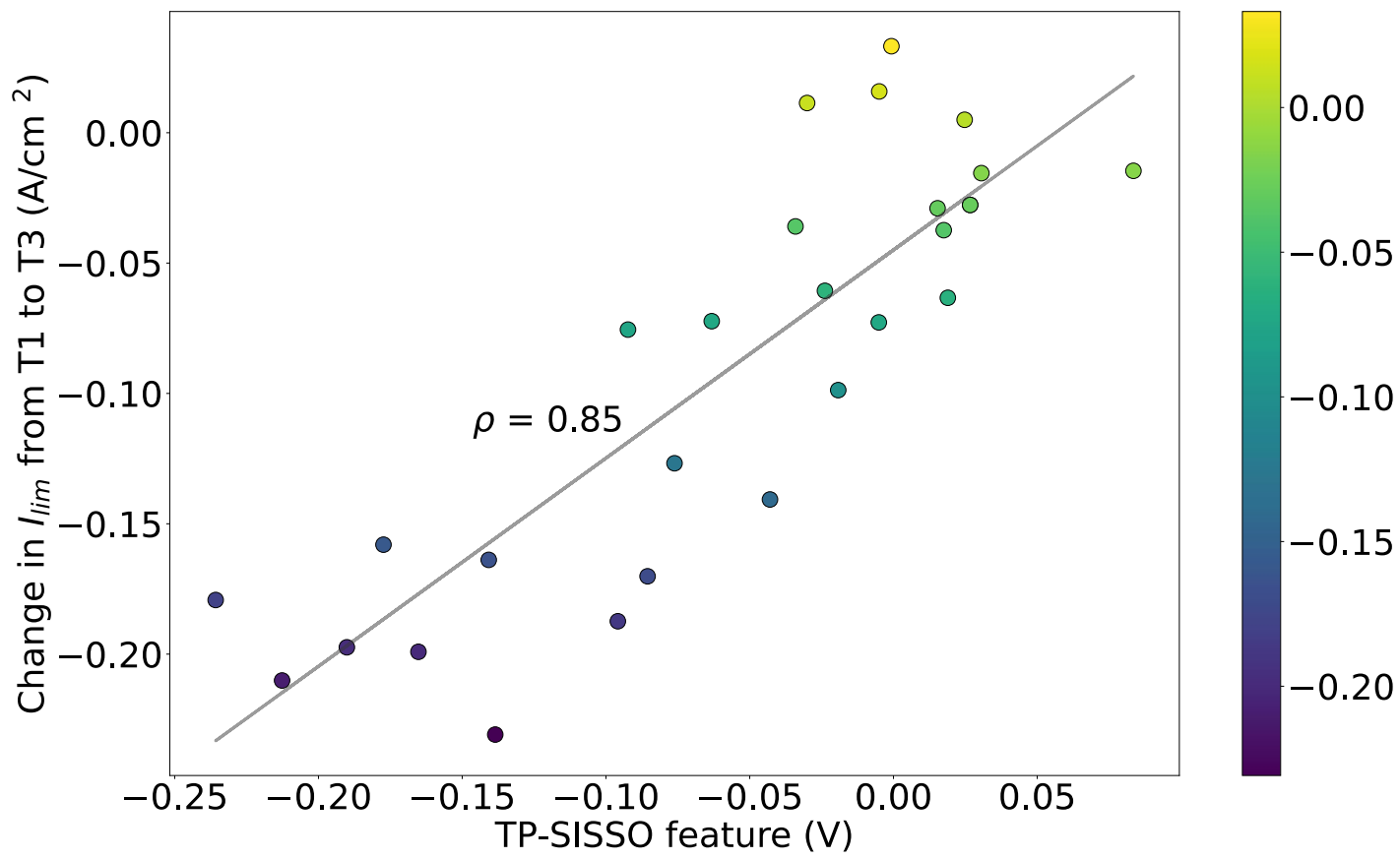

**Supplementary Figure 12** The relationship between the TP-SISSO feature extracted from the ΔV/I curve and the change in $I_{lim}$ from the T1 stage to T3 stage of Dataset 1. The change in $I_{lim}$ from the T1 stage to T3 stage is calculated by subtracting the $I_{lim}$ value of T1 stage from that of T3 stage.

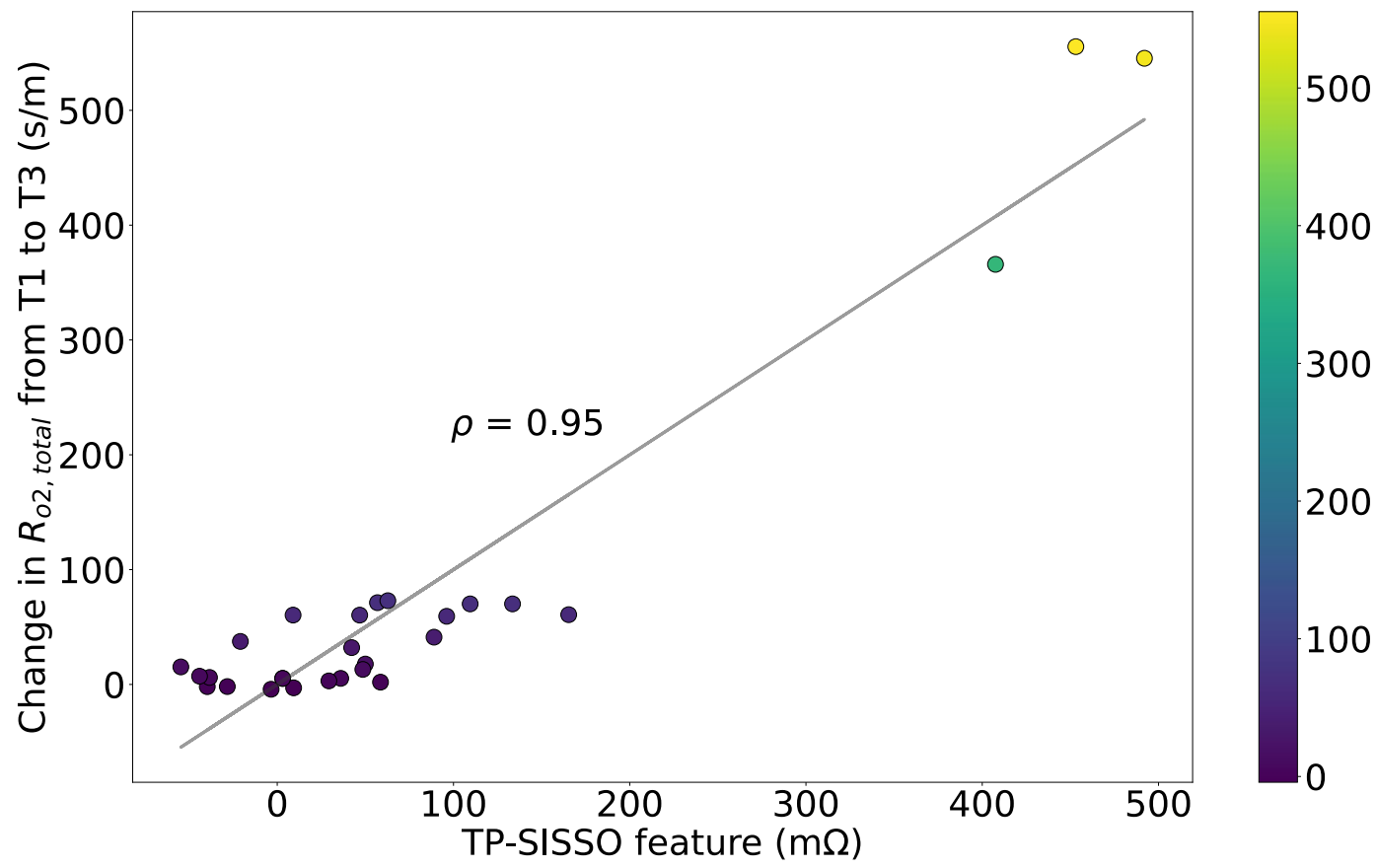

(a) The relationship between the TP-SISSO feature extracted from the ΔRe/f curve and the change in $R_{o2,total}$ from the T1 stage to T3 stage. The change in $R_{o2,total}$ from the T1 stage to T3 stage is calculated by subtracting the $R_{o2,total}$ value of T1 stage from that of T3 stage.

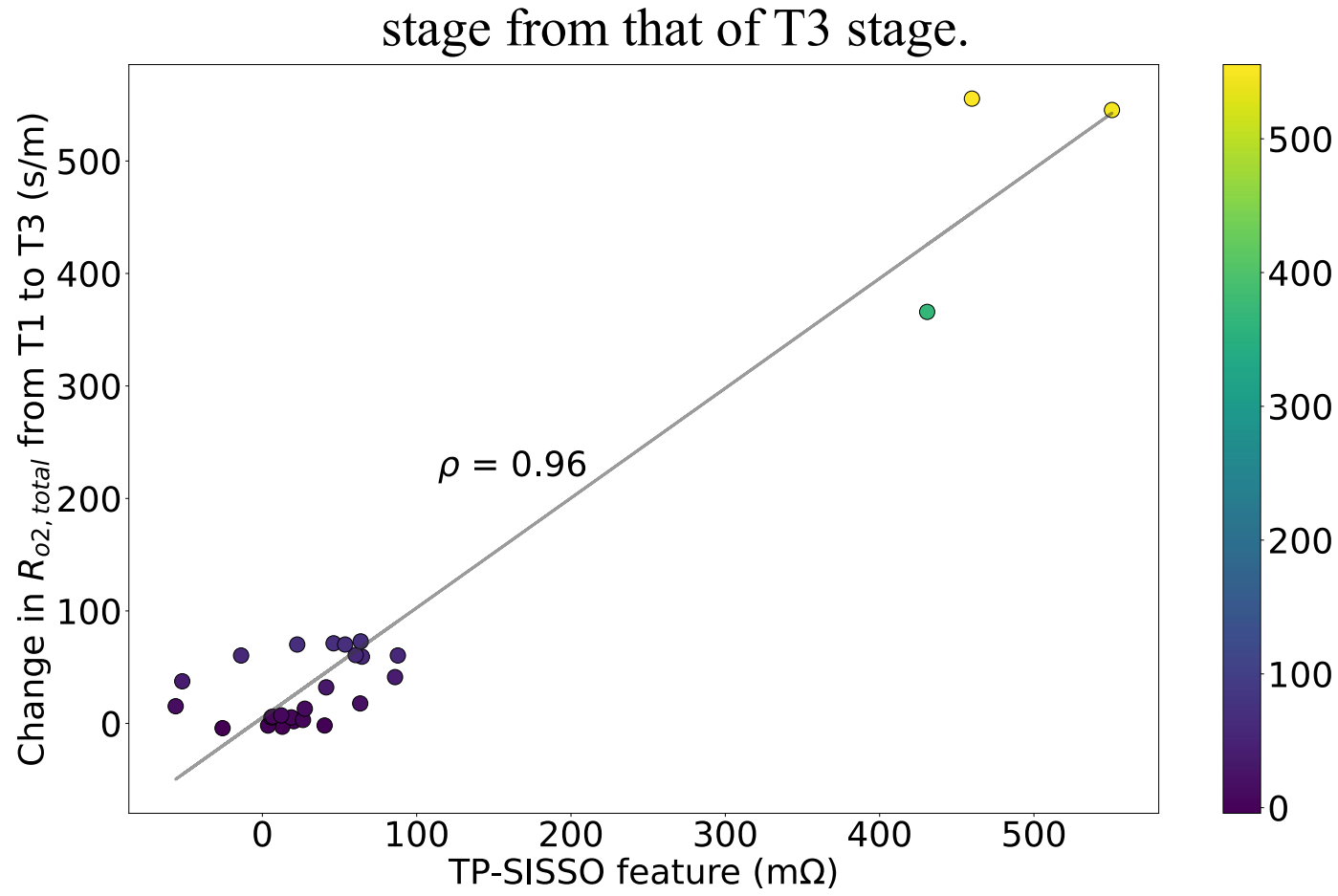

(b) The relationship between the TP-SISSO feature extracted from the ΔIm/f curve and the change in $R_{o2,total}$ from the T1 stage to T3 stage. The change in $R_{o2,total}$ from the T1 stage to T3 stage is calculated by subtracting the $R_{o2,total}$ value of T1 stage from that of T3 stage.

**Supplementary Figure 13** The relationship between the TP-SISSO feature extracted from the EIS and the change in $R_{o2,total}$ from the T1 stage to T3 stage of Dataset 1.

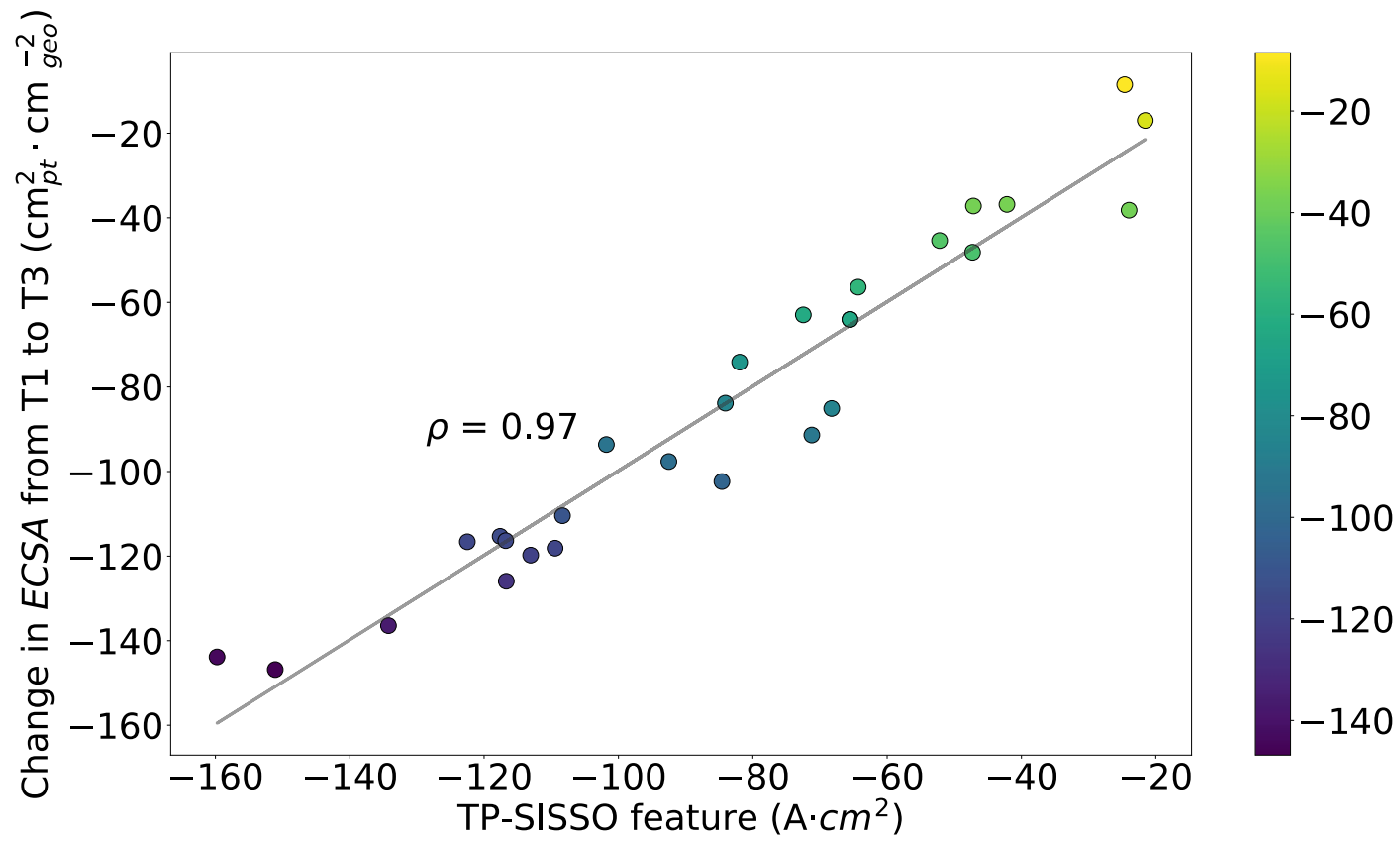

**Supplementary Figure 14** The relationship between the TP-SISSO feature extracted from the ΔI/V curve and the change in $ECSA$ from the T1 stage to T3 stage of Dataset 1. The change in $ECSA$ from the T1 stage to T3 stage is calculated by subtracting the $ECSA$ value of T1 stage from that of T3 stage.

## Supplementary Note 7

Based on the candidate two-point features, this paper adopts the SISSO algorithm [4] to extract the TP-SISSO feature. This paper adopts the *SissoModel* function in the *TorchSisso* Python library [5]. Specifically, all the obtained candidate two-point features are input into the *SissoModel* function, which automatically outputs the TP-SISSO feature calculation formula composed of $k$ candidate two-point features. The hyperparameters of the *SissoModel* function include *operators*, *n_expansion*, *use_gpu*, and $k$. In this paper, the hyperparameters are set to *operators*=[+, -], *n_expansion*=2, *use_gpu*=True, and $k$=6.

## Supplementary Note 8

As shown in **Supplementary Figure 15**, in Dataset 1, the input of the $I_{lim}$ prediction model is the TP-SISSO feature extracted from the ΔV/I curve, and the output is the predicted value of the change in $I_{lim}$ from the T1 stage to T3 stage. The inputs of the $R_{o2,total}$ prediction model are two TP-SISSO features extracted from the ΔRe/f curve and ΔIm/f curve, and the output is the predicted value of the change in $R_{o2,total}$ from the T1 stage to T3 stage. The input of the $ECSA$ prediction model is the TP-SISSO feature extracted from the ΔI/V curve, and the output is the predicted value of the change in $ECSA$ from the T1 stage to T3 stage.

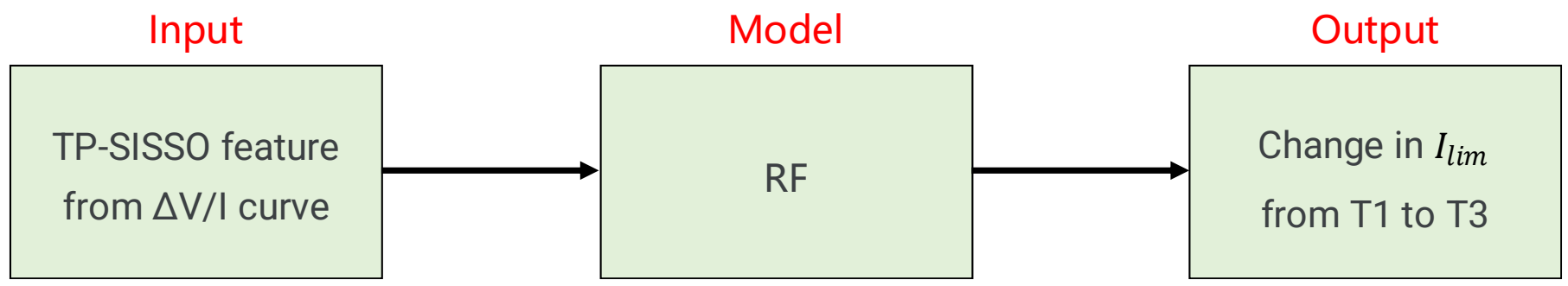

(a) Prediction method for $I_{lim}$.

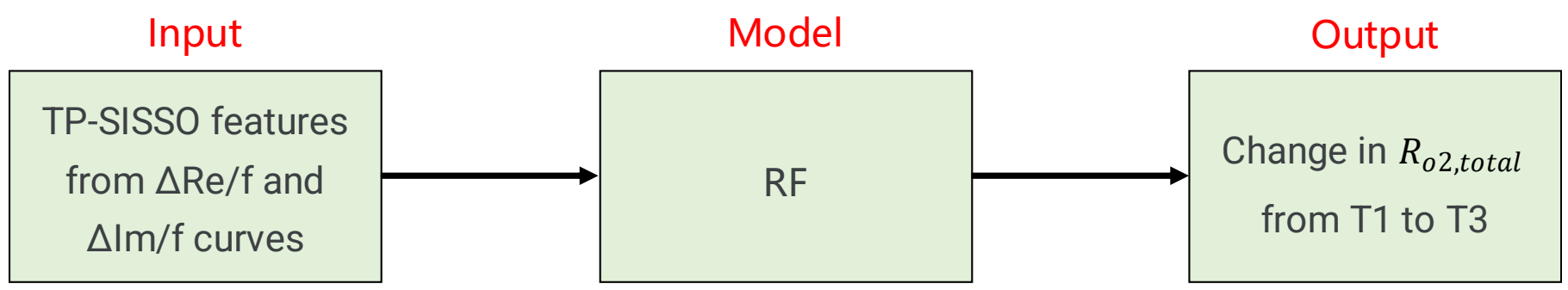

(b) Prediction method for $R_{o2,total}$.

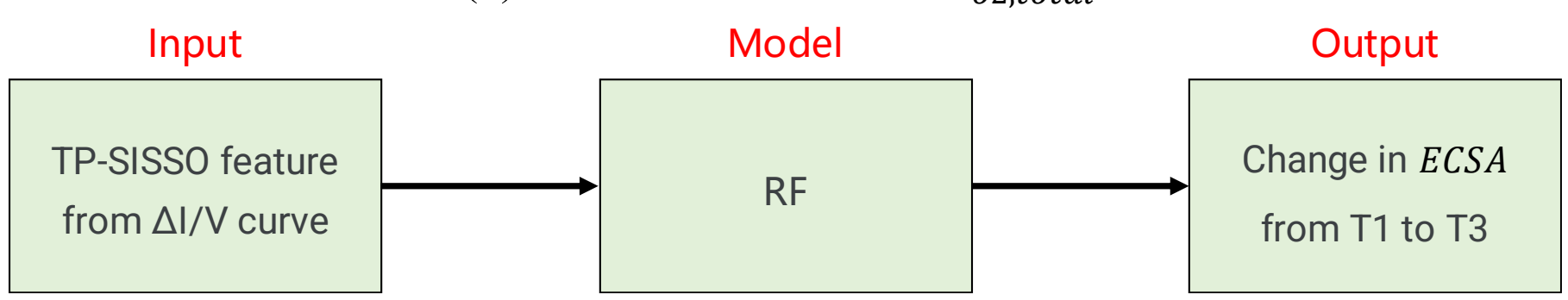

(c) Prediction method for $ECSA$.

**Supplementary Figure 15** Prediction methods for different aging indicators in the T3 stage of Dataset 1.

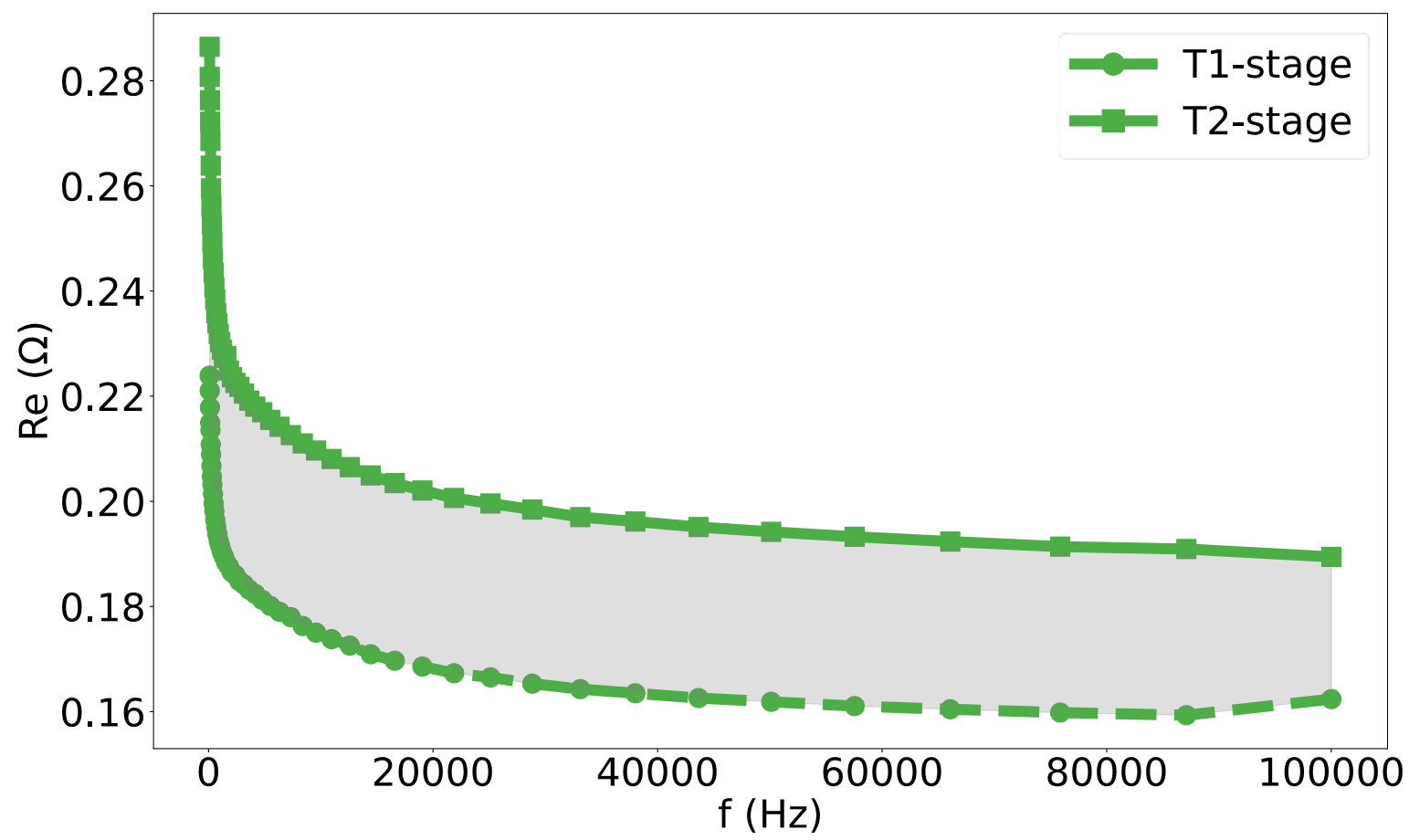

(a) Calculation method of ΔRe/f curve.

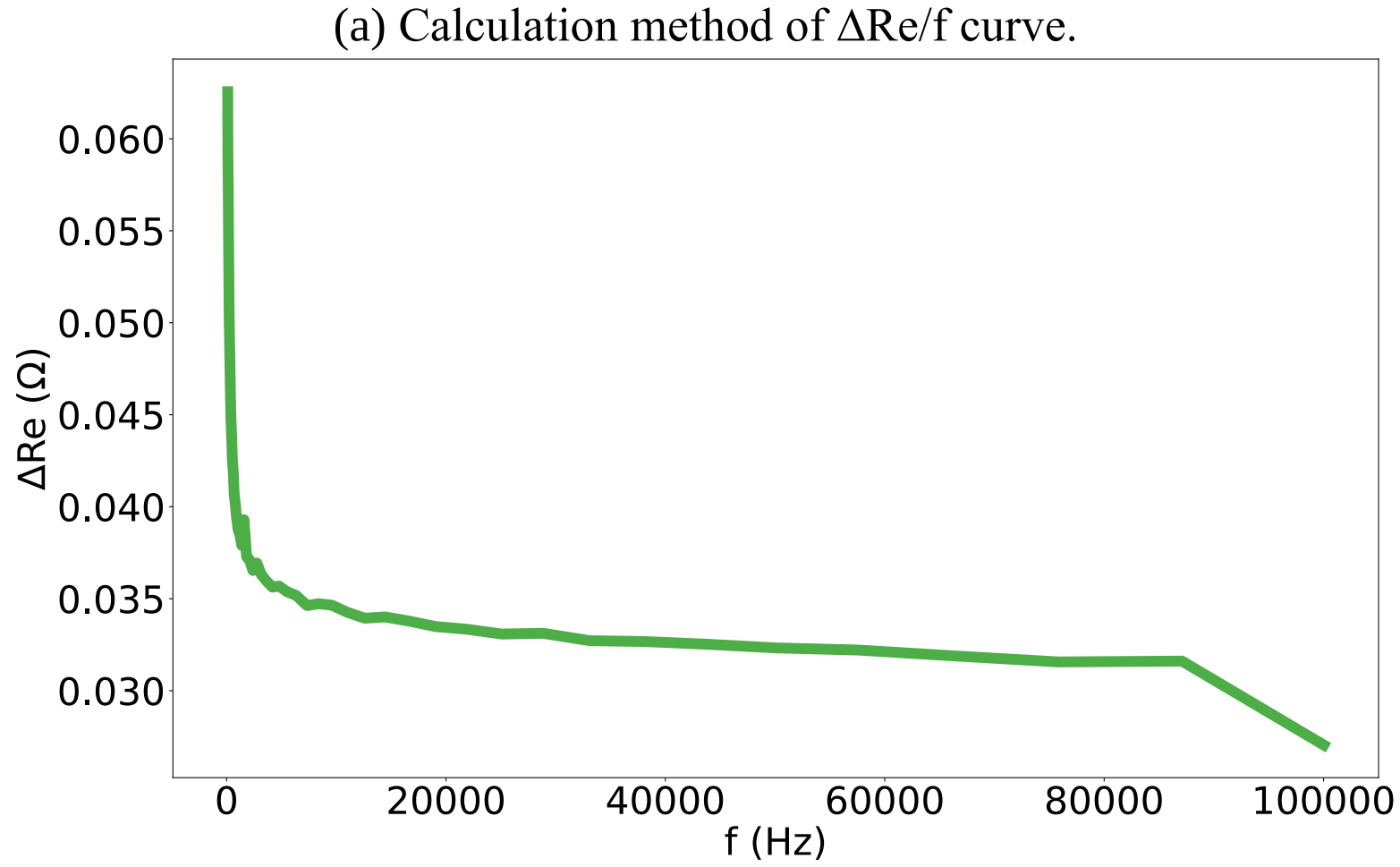

(b) Calculation results of ΔRe/f curve.

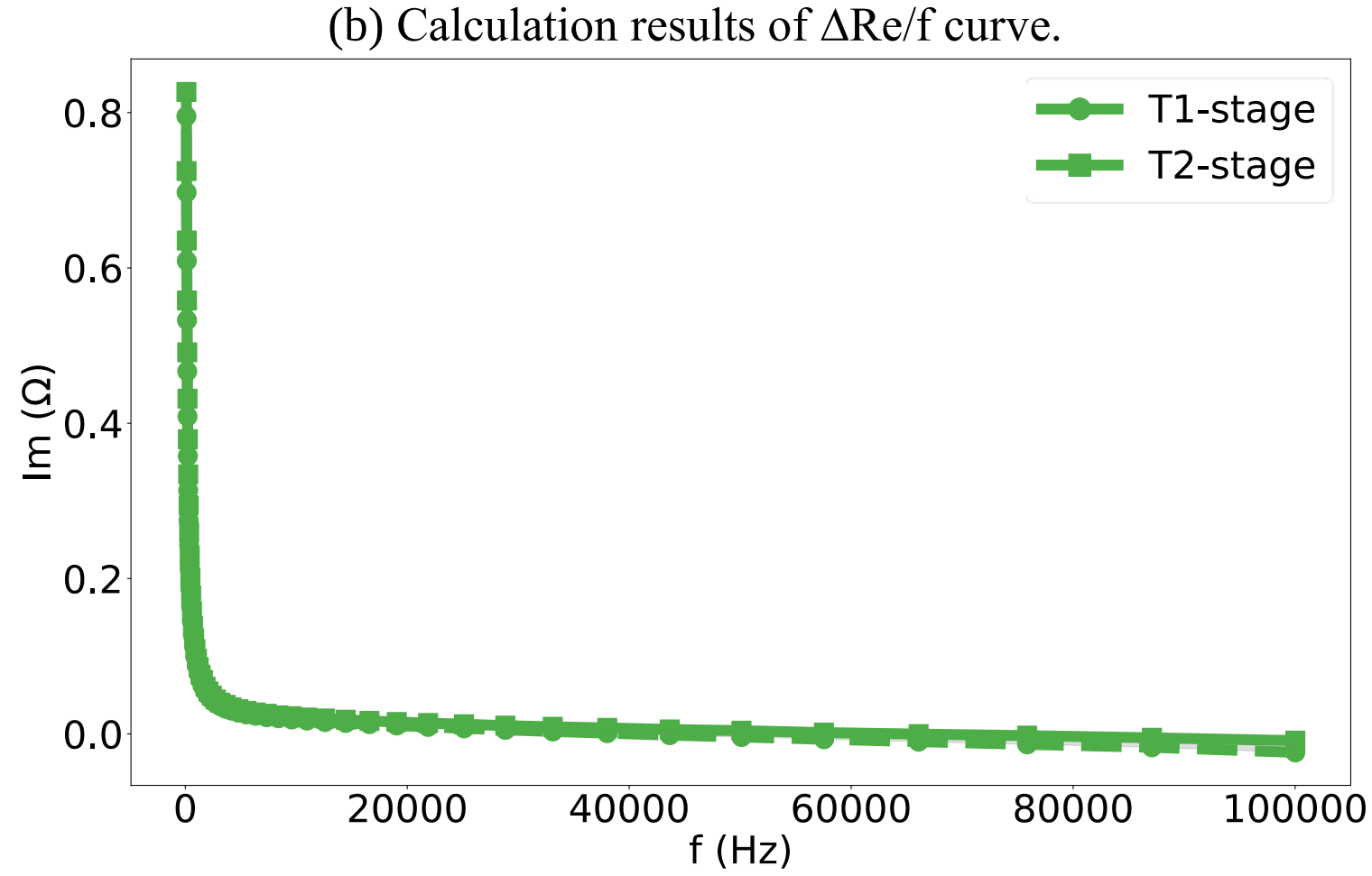

(c) Calculation method of ΔIm/f curve.

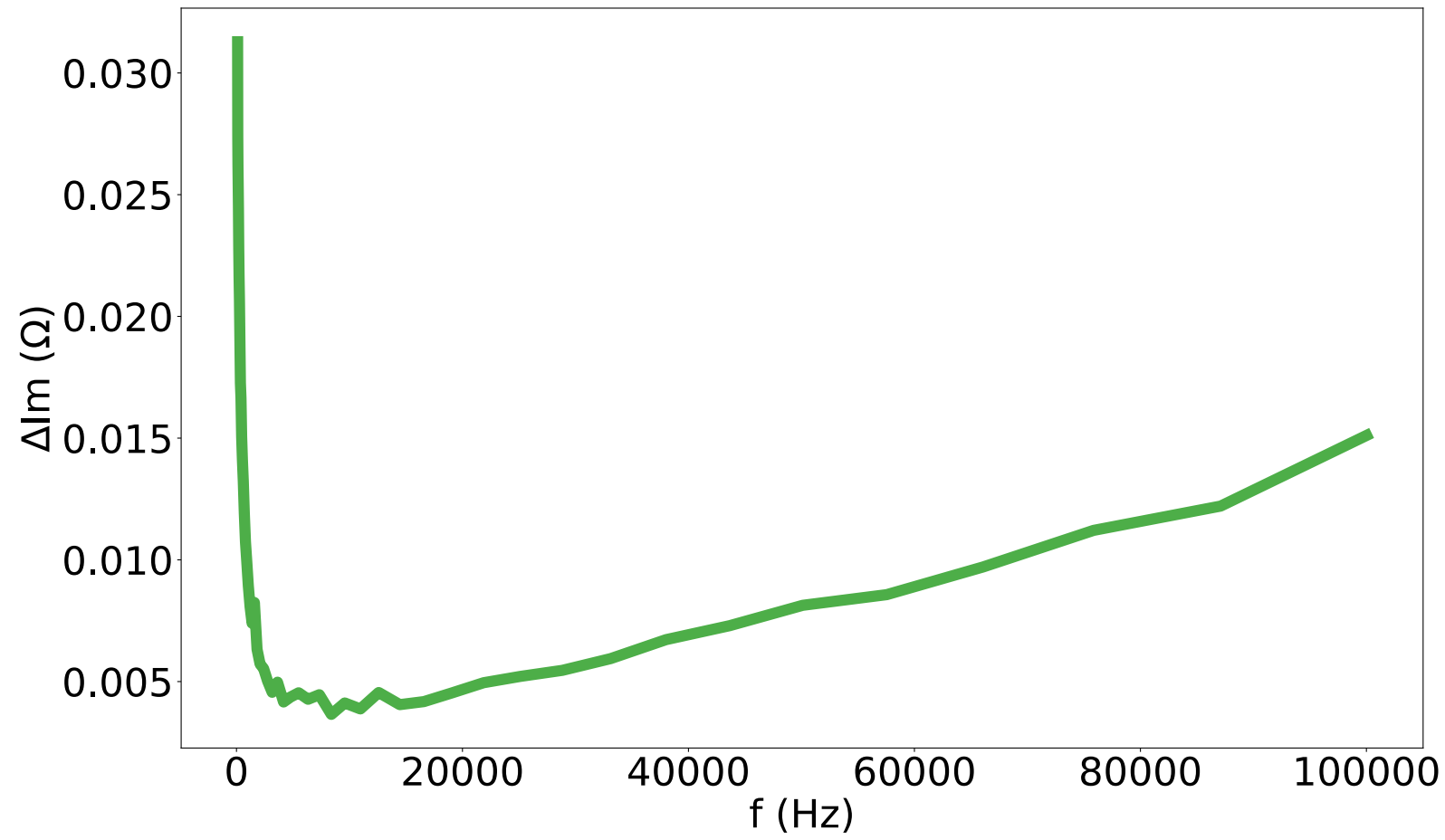

(d) Calculation results of ΔIm/f curve.

**Supplementary Figure 16** Calculation of the ΔRe/f curve and the ΔIm/f curve from the representative capacitor (numbered as ES10C4 in Dataset 3).

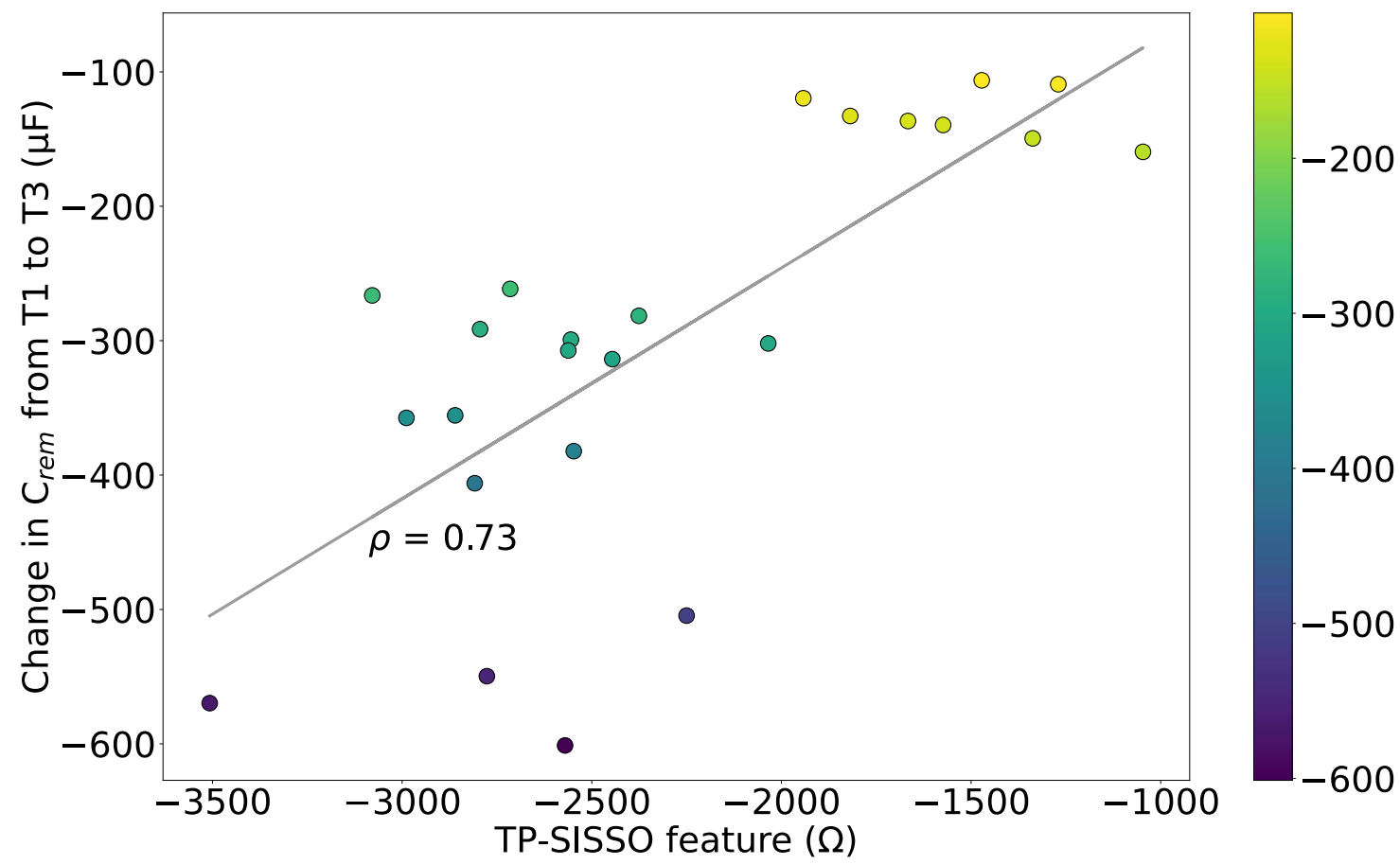

(a) The relationship between the TP-SISSO feature extracted from the ΔRe/f curve and the change in $C_{rem}$ from the T1 stage to T3 stage. The change in $C_{rem}$ from the T1 stage to T3 stage is calculated by subtracting the $C_{rem}$ value of T1 stage from that of T3 stage.

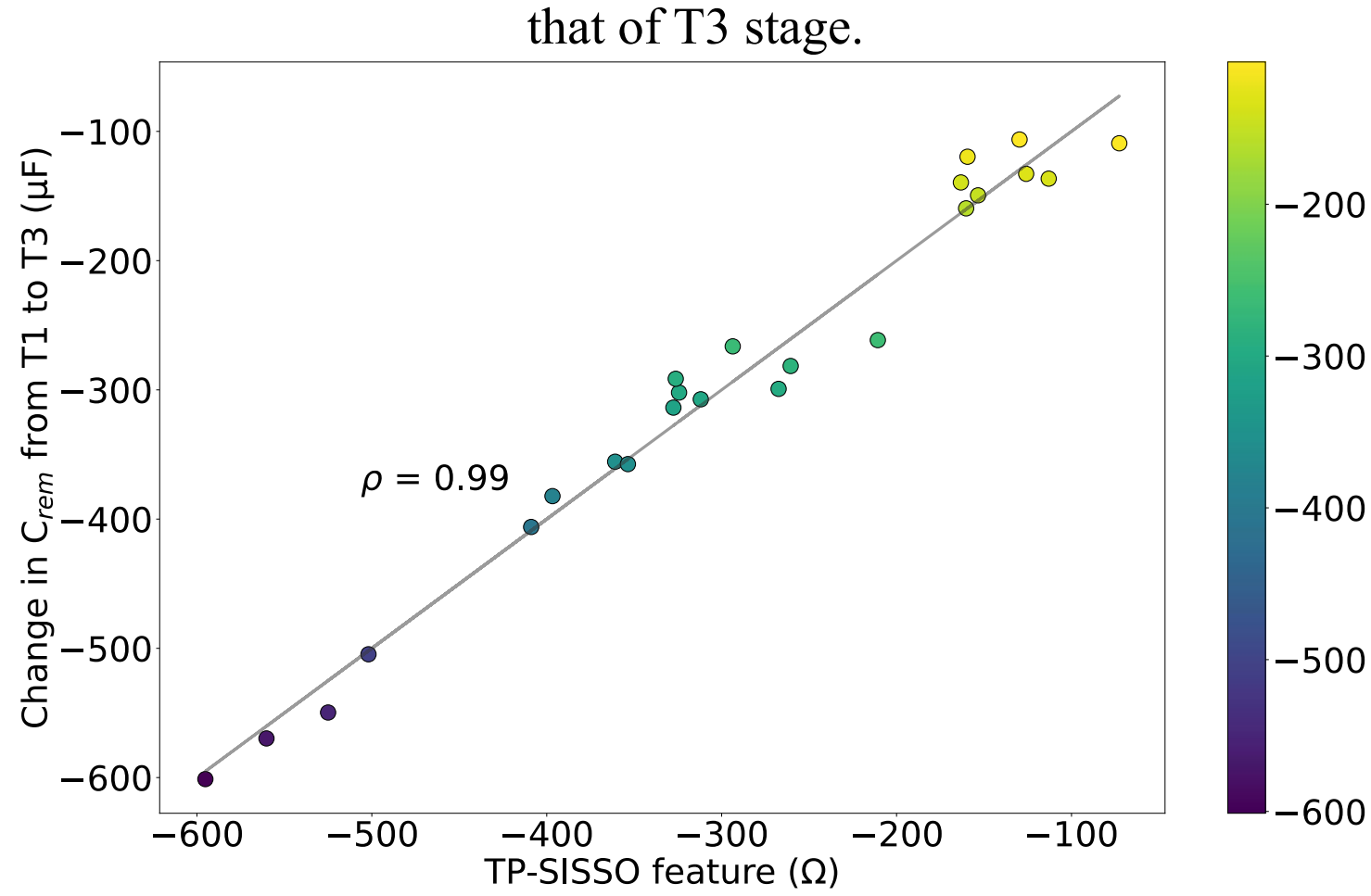

(b) The relationship between the TP-SISSO feature extracted from the ΔIm/f curve and the change in $C_{rem}$ from the T1 stage to T3 stage. The change in $C_{rem}$ from the T1 stage to T3 stage is calculated by subtracting the $C_{rem}$ value of T1 stage from that of T3 stage.

**Supplementary Figure 17** The relationship between the TP-SISSO feature extracted from the EIS and the change in $C_{rem}$ from the T1 stage to T3 stage of Dataset 3.